\documentclass[12pt]{article}
\usepackage{amsfonts}
\usepackage[table]{xcolor}

\usepackage{amsbsy}

\usepackage{caption}
\usepackage{amsmath}
\usepackage{algorithm}
\usepackage{algorithmic}

\usepackage{amssymb}

\usepackage{graphicx}
\usepackage{subcaption}

\usepackage{amsthm}

\usepackage{rotating}

\newcommand{\blind}{1}

\usepackage{graphics}

\usepackage{epstopdf}
\usepackage{hyperref}
\hypersetup{colorlinks,
citecolor=blue
}

\usepackage{pdfsync}

\usepackage{caption}

\usepackage{bm}
 \usepackage{graphicx,color}
\usepackage{setspace}
\usepackage{natbib}  
\makeatletter 

\def\singlespace{\def\baselinestretch{1}\@normalsize}

\@addtoreset{equation}{section}

\newtheorem{theorem}{Theorem}

\newtheorem{condition}{Condition}

\newcommand{\distas}[1]{\mathbin{\overset{#1}{\kern\z@\sim}}}%

\def\hat{\widehat}

\def\tilde{\widetilde}

\def\bX{{\bf X}}

\newcommand{\T}{\!\top\!}

\newcommand{\footremember}[2]{%
	\footnote{#2}
	\newcounter{#1}
	\setcounter{#1}{\value{footnote}}%
}
\newcommand{\footrecall}[1]{%
	\footnotemark[\value{#1}]%
}

\date{}

\begin{document}

\if1\blind
{ \title{\bf Multicategory Angle-based Learning for Estimating Optimal Dynamic Treatment Regimes with Censored Data}
  \author{Fei Xue\footremember{first author}{These authors contributed equally to this work.}\\
   \normalsize Department of Biostatistics, Epidemiology and Informatics, University of Pennsylvania,\\
	Yanqing Zhang\footrecall{first author}\\
  \normalsize Department of Statistics, Yunnan University, \\
	Wenzhuo Zhou\footrecall{first author}\\
 \normalsize  Department of Statistics, University of Illinois at Urbana-Champaign, \\
	Haoda Fu \\
  \normalsize Eli Lilly, Clinical Research Department,\\
	Annie Qu\thanks{
    The author is supported by \textit{National Science Foundation Grants DMS-1821198}.}\hspace{.2cm}\\
   \normalsize Department of Statistics, University of Illinois at Urbana-Champaign\\
   }
    \maketitle
    \newpage
} \fi

\if0\blind
{
  \bigskip
  \bigskip
  \bigskip
  \begin{center}
    {\LARGE\bf Multicategory Angle-based Learning for Estimating Optimal Dynamic Treatment Regimes with Censored Data}
\end{center}
  \medskip
} \fi


\begin{abstract}

An optimal dynamic treatment regime (DTR) consists of a sequence of decision rules in maximizing long-term benefits, which is applicable for chronic diseases such as HIV infection or cancer.
	In this paper, we develop a novel angle-based approach to search the optimal DTR under a multicategory treatment framework for survival data.
	The proposed method targets maximization the conditional survival function of patients following a DTR.
In contrast to most existing approaches which are designed to maximize the expected survival time under a binary treatment framework,
the proposed method solves the multicategory treatment problem given multiple stages for censored data.
	Specifically, the proposed method obtains the optimal DTR via integrating estimations of decision rules at multiple stages into a single multicategory classification algorithm without imposing additional constraints, which is also more computationally efficient and robust.
In theory, we establish Fisher consistency of the proposed method under regularity conditions. Our numerical studies show that the proposed method outperforms competing  methods in terms of maximizing the conditional survival function. We apply the proposed method to two real datasets: Framingham heart study data and acquired immunodeficiency syndrome (AIDS) clinical data.

\end{abstract}


\newpage

\section{Introduction}\label{sec:1}

			Precision medicine tailored for individuals has become an important strategy in treating chronic diseases and conditions of patients. Dynamic treatment regimes (DTR) play a central role in precision medicine, such as
		recommendation of optimal treatments to individual patients according to patients' previous treatments and medical histories. A {DTR} is also called an adaptive intervention \citep{collins2004conceptual}, or adaptive strategy \citep{lavori2000design} under other contexts. In practice, an effective treatment strategy should not focus on short-term benefits, but aim for the most favorable long-term benefits. Consequently, the goal is to seek an optimal DTR, which is defined as a sequence of decision functions, to maximize the expected long-term benefits \citep{murphy2003optimal,murphy2005experimental}.
		
		For chronic diseases such as cancer or HIV infection, survival time is often the outcome of interest, and  developing  an optimal DTR is critical under the survival  data framework. Although many  existing works have made important contributions to the estimation of the optimal DTR \citep{watkins1992q,murphy2003optimal,murphy2005experimental,chakraborty2010inference,moodie2010estimating,robins2008estimation,zhang2013robust,zhao2015new,zhu2017greedy,robins2000marginal,blatt2004learning}, very few works focus on maximizing survival probability. Yet, maximizing survival probability is also vital for patients. There are two main challenges in estimating the optimal DTR for censored data. The first one is that there could be
		 a lack of treatment and covariate information from patients  in follow-up stages due to censoring.
		The second one is that the true survival time might be unknown for patients who are still alive at the {censoring time}
		\citep{goldberg2012q}. Methods proposed by	\cite{zhao2011reinforcement,jiang2017estimation,bai2017optimal,huling2018subgroup,zhao2014doubly,hager2018optimal} only {focus on} one or two {decision points (stages)}, {while the censored Q-learning algorithm \citep{goldberg2012q} and the stabilized O-learning approach \citep{zhao2018constructing} are able to deal with settings including more than two stages.}

		{Nevertheless}, none of the aforementioned methods can handle multicategory treatment scenarios when constructing a {DTR}. Two common approaches to solving multicategory classification problems are one-versus-one and one-versus-rest approaches \citep{allwein2000reducing}, which both apply sequential binary classifiers. However, the sequential binary classifier only {yields} a sub-optimal result in some cases \citep{zhang2013multicategory}. Another approach to consider multiple treatment choices simultaneously
		is to  estimate a classification function vector with a dimension
		determined by the number of category \citep{crammer2001algorithmic, hastie2009multi, liu2011reinforced, vapnik1998statistical}. The corresponding decision rule assigns a subject to the category with the largest estimated value in the classification function vector for this subject.
		Typically, a sum-to-zero constraint, requiring the sum of values from  the classification function vector to be zero, is used to reduce the parameter space for desirable theoretical properties \citep{lee2004multicategory,liu2011reinforced}. However,  additional  computational cost is needed to solve the constrained optimization problem which could  be  inefficient \citep{zhang2014multicategory}.
		
		To overcome these drawbacks,  \cite{zhang2014multicategory} proposed an
		angle-based large-margin classifier, which can significantly  reduce the computational cost without additional constraints, and outperforms other standard classification methods. To adapt this technique to precision medicine, \cite{zhang2018multicategory} formulated  a weighted angle-based method to develop an  individualized treatment rule with  multiple treatment {choices}. \cite{tao2017adaptive} proposed an adaptive contrast weighted learning method to identify the optimal DTR in a multicategory treatment setting. However, those approaches are neither  applicable  for multiple  decision points nor can they handle censored outcomes.

		In this paper, we propose a new angle-based weighted approach for estimation of the optimal DTR  by maximizing the conditional survival probability under a multicategory treatment framework. Specifically, we  propose a weighted 
		Kaplan-Meier (KM) estimator to estimate the survival function {under a DTR through the inverse of the treatment probability at each stage.} 
	One key idea is to decompose the survival function at a given time-point  
			to a product  of survival probabilities at time-points before the given time. 
			An advantage of this decomposition is that we only need  to consider treatments received at stages before each of these previous time-points,
			since the survival probability at a certain time cannot be affected by  treatments received later than that time. In this way, we can incorporate all available observations from  patients into a weighted KM estimator, even though some patients may lack treatment and covariate information at certain stages due to censoring.
		In addition, we adopt an angle-based classifier to consider multiple potential treatment choices
		 simultaneously and
		avoid the sum-to-zero constraint, which could be  computationally restrictive in optimization.

		Moreover, we propose to estimate  {decision rules for all stages simultaneously} by maximizing the proposed objective function over a class of treatment regimes. In other words, our method integrates the estimations of decision rules at  multiple stages as one joint weighted multicategory classification problem. This simultaneous procedure enables our approach to be more  robust by circumventing potential model-misspecification problems arising in the Q-learning approaches, which recursively fit posited regression models based on the estimations of regression models at future stages
		\citep{zhao2015new,zhou2017augmented}.

		The main contributions of this paper {can be summarized as follows.} First, to the best of our knowledge, this is the first work which directly estimates  the optimal DTR 
		under a multicategory treatment framework {for} censored data.
		Second, the proposed method  maximizes  the survival probability in searching an optimal DTR with more than two decision stages. This has a profound impact in practice, since there is a high demand from long-term treatment management for making multistage decisions in  treating  chronic diseases.
Third, the proposed method transforms DTR estimation problems at multiple stages into a joint multicategory classification problem and solves the optimization problem without imposing additional  constraints, which improves computational efficiency and avoids potential overfitting problems.
		In theory,  we establish Fisher consistency 
		of the proposed method, which has not previously been established for the estimation of DTR under the  multicategory treatment framework.

		The remainder of the paper is organized as follows. In Section \ref{sec:2}, we introduce the background and notation of the {DTR} and survival analysis. In Section \ref{sec:3}, we propose an angle-based weighted Kaplan-Meier estimator for the conditional survival function {under} a DTR. Based on this estimator, we propose a novel estimation approach for the optimal DTR under the multicategory treatment framework. The {Fisher} consistency of the proposed method is established in Section \ref{sec:4}. The computation algorithm is provided in Section \ref{sec:5}. In Section \ref{sec:6}, we present empirical comparisons of the proposed method with {the} censored Q-learning and subgroup identification method.
		Section \ref{sec:7} illustrates the application of the proposed method for the Framingham Heart data and AIDS clinical data. Finally, we conclude with discussion in Section \ref{sec:8}.

\section{Background and notations}\label{sec:2}

In this section, we introduce background and notations for the standard dynamic treatment regime setting and survival analysis. We assume that the observed data are $$\{(\bm{X}_{i1},A_{i1},\ldots,\bm{X}_{im(Y_i)},A_{im(Y_i)},Y_i)\}^n_{i=1},$$
containing  $n$ independent, identically distributed samples of
 $(\bm{X}_{1},A_{1},\ldots,\bm{X}_{m(Y)}, A_{m(Y)}, \linebreak Y)$, where $\bm{X}_{m}\in\mathbb{R}^p$ denotes the covariates
 information collected between the $(m-1)$-th and $m$-th stages, $A_{m} \in \mathcal{A}$ denotes the treatment assigned at the $m$-th stage, $Y = T \wedge C$ denotes the observed survival time, and $T$ and $C$ correspond to the survival time and censoring time, respectively.
Here, $m(t)$ represents the index of the stage where the time-point $t$ belongs, implying  that
a subject is either censored, or a failure event occurs at the stage $m(Y)$.
In addition,  we let $\Delta=I(T \le C)$ be an indicator for the occurrence of the failure event at or before the censoring time, and let $\widetilde{\bm{A}}_{m}=(A_{1}, \dots, A_{m})$ and $\widetilde{\bm{X}}_{m}=(\bm{X}_{1}, \dots, \bm{X}_{m})$ be  longitudinal combinations of treatments and covariates from the first stage to the $m$-th stage, respectively.  Let $\bm{H}_{m}=(\widetilde{\bm{X}}_{m},\widetilde{\bm{A}}_{m-1})$  denote accrued information up to the $m$-th stage in which $\bm{H}_{1}=\bm{X}_{1}$. Throughout this paper, we make the non-informative censoring assumption that the censoring time $C$ is independent of survival time, covariates and treatments; that is, the censoring is random and non-informative.

For a given time-point $t_g$, our goal is to find a sequence of decision rules $\widetilde{\bm{\mathcal{D}}}_{m_g}(\bm{H}_{m_g})=\{\mathcal{D}_1(\bm{H}_1), \dots, \mathcal{D}_{m_g}(\bm{H}_{m_g})\}$ under which  the survival function at $t_g$
\begin{eqnarray}\label{SProb}
\hspace{-8mm}S^{\widetilde{\bm{\mathcal{D}}}_{m_g}}(t_g)&=&E_{1}\left[E_{2}^{\widetilde{\bm{\mathcal{D}}}_{1}}\left[\cdots E_{m_g}^{\widetilde{\bm{\mathcal{D}}}_{m_g-1}}\left[P\{T>t_g| \tilde{\bm{X}}_{m_g}, \widetilde{\bm{A}}_{m_g} = \widetilde{\bm{\mathcal{D}}}_{m_g}(\bm{H}_{m_g})\}\right]\cdots\right]\right]\notag\\
&=&E_{1}\left[E_{2}^{\widetilde{\bm{\mathcal{D}}}_{1}}\left[\cdots E_{m_g}^{\widetilde{\bm{\mathcal{D}}}_{m_g-1}}\left[E\{I(T>t_g)| \tilde{\bm{X}}_{m_g}, \widetilde{\bm{A}}_{m_g} = \widetilde{\bm{\mathcal{D}}}_{m_g}(\bm{H}_{m_g})\}\right]\cdots\right]\right],
\end{eqnarray}
is  maximized, 
where  $E_1$ and $E_{m}^{\widetilde{\bm{\mathcal{D}}}_{m-1}}$ represent expectations with respect to $\bm{X}_1$ and $\{\bm{X}_m \mid \tilde{\bm{X}}_{m-1}, \widetilde{\bm{A}}_{m-1} = \widetilde{\bm{\mathcal{D}}}_{m-1}(\bm{H}_{m-1})\}$, respectively,   for $2\le m \le m_g$. Here, $m_g=m(t_{g-1})$ since the survival probability at time $t_g$ is assumed to only be affected by treatments at or  before time $t_{g-1}$.

\section{Angle-based weighted Kaplan-Meier method}\label{sec:3}
\subsection{Survival function under a sequence of decision rules}

To estimate an optimal {DTR} which maximizes the $S^{{\widetilde{\bm{\mathcal{D}}}}_{m_g}}(t_g)$ in (\ref{SProb}), we first derive a weighted Kaplan-Meier estimator for the survival function $S^{{\widetilde{\bm{\mathcal{D}}}}_{m_g}}(t_g)$ as follows. Let $E^{\widetilde{\bm{\mathcal{D}}}_{m_g}}$ denote the expectation with respect to distribution $ \mathcal{P}^{\widetilde{\bm{\mathcal{D}}}_{m_g}}$, which is the conditional distribution of $(T, \bm{H}_{m(t)}, A_{m(t)})$ under $ \widetilde{\bm{A}}_{m(t)} = \widetilde{\bm{\mathcal{D}}}_{m(t)}(\bm{H}_{m(t)})$ with probability density function
\begin{eqnarray}\label{dis}
p(T|\bm{H}_{m_g},A_{m_g})I[\widetilde{\bm{A}}_{m_g}=\widetilde{\bm{\mathcal{D}}}_{m_g}(\bm{H}_{m_g})] \prod_{j=1}^{m_g}p(\bm{X}_{j}|\bm{H}_{j-1},A_{j-1}),
\end{eqnarray}
where $p(\bm{X}_1|\bm{H}_{0},A_{0})=p(\bm{X}_1)$.

Since the censoring is random and able to occur at any time-point before $t_g$, we adopt the Kaplan-Meier estimator to decompose the survival probability $S^{{\widetilde{\bm{\mathcal{D}}}}_{m_g}}(t_g)$ in terms of survival probabilities at time-points before $t_g$, and estimate $S^{{\widetilde{\bm{\mathcal{D}}}}_{m_g}}(t_g)$ as:
\begin{eqnarray}
S^{\widetilde{\bm{\mathcal{D}}}_{m_g}}(t_g)=E^{\widetilde{\bm{\mathcal{D}}}_{m_g}}\left[I(T>t_g)\right]=\prod^{g}_{s=1}q^{\widetilde{\bm{\mathcal{D}}}_{m_g}}(s),
\end{eqnarray}
where
\begin{eqnarray}
q^{\widetilde{\bm{\mathcal{D}}}_{m_g}}(s)=\frac{ E^{\widetilde{\bm{\mathcal{D}}}_{m_g}}\left[I(T \ge t_{s+1})\right]}{E^{\widetilde{\bm{\mathcal{D}}}_{m_g}}\left[I(T\ge t_{s})\right]}.
\end{eqnarray}

Note that the occurrence of a failure event before or at a specific time $t_s$  can only be affected by the treatments and covariates before $t_s$. We assume that
\begin{equation}
I(T=t_s \; \text{or} \; T\ge t_s)
\perp \!\!\! \perp Z
\mid \{ \tilde{\bm{X}}_{m(t_{s-1})}, \widetilde{\bm{A}}_{m(t_{s-1})} = \widetilde{\bm{\mathcal{D}}}_{m(t_{s-1})}\big(\bm{H}_{m(t_{s-1})}\big) \},
\end{equation}
where $Z$ represents $\bm{X}_{m}$ or $\bm{A}_{m} = \bm{\mathcal{D}}_{m}(\bm{H}_{m})$ for $m>m(t_{s-1})$.  Under the above assumption, we can ignore the influence of the treatments after the {$m(t_{s-1})$-th stage}, and the corresponding modified $q^{\widetilde{\bm{\mathcal{D}}}_{m_g}}(s)$ is
\begin{eqnarray}
q^{\widetilde{\bm{\mathcal{D}}}_{m_g}}(s)=1-\frac{ E^{\widetilde{\bm{\mathcal{D}}}_{m_g}}\left[I(T=t_s)\right]}{E^{\widetilde{\bm{\mathcal{D}}}_{m_g}}\left[I(T\ge t_s)\right]}=1-\frac{ E^{\widetilde{\bm{\mathcal{D}}}_{m(t_{s-1})}}\left[I(T=t_s)\right]}{E^{\widetilde{\bm{\mathcal{D}}}_{m(t_{s-1})}}\left[I(T\ge t_s)\right]}.\label{qequation1}
\end{eqnarray}

However, samples from the distribution $\mathcal{P}^{\widetilde{\bm{\mathcal{D}}}_{m_g}}$ with {the density in (\ref{dis})} are generally not observable, which implies that $E^{\widetilde{\bm{\mathcal{D}}}_{m(t_{s-1})}}\left[I(T\ge t_s)\right]$ cannot be calculated based on the observed data directly. To estimate $q^{\widetilde{\bm{\mathcal{D}}}_{m_g}}(s)$, we
convert the expectation $E^{\widetilde{\bm{\mathcal{D}}}_{m(t_{s-1})}}$ to
an unconditional expectation using the Radon-Nikodym theorem. That is, 
\begin{equation}\label{eq:main}
E^{\widetilde{\bm{\mathcal{D}}}_{m(t_{s-1})}}\left[I(T\ge t_s)\right]=\int I(T\ge t_s) d \mathcal{P}^{\widetilde{\bm{\mathcal{D}}}_{m(t_{s-1})}}= \int I(T\ge t_s) \frac{d \mathcal{P}^{\widetilde{\bm{\mathcal{D}}}_{m(t_{s-1})}}}{d \mathcal{P}_{m(t_{s-1})}} d \mathcal{P}_{m(t_{s-1})},
\end{equation}
where $\mathcal{P}_{m(t_{s-1})}$ denotes the unconditional distribution of $(T, \bm{H}_{m(t_{s-1})}, A_{m(t_{s-1})} )$ with the following probability density function
$$p(T|\bm{H}_{m(t_{s-1})},A_{m(t_{s-1})})\prod_{j=1}^{m(t_{s-1})}p(A_{j}|\bm{H}_{j})p(\bm{X}_{j}|\bm{H}_{j-1},A_{j-1}).$$ Since
\begin{eqnarray*}
	\frac{d \mathcal{P}^{\widetilde{\bm{\mathcal{D}}}_{m(t_{s-1})}}}{d \mathcal{P}_{m(t_{s-1})}}
	=\frac{I[\widetilde{\bm{A}}_{m(t_{s-1})}=\widetilde{\bm{\mathcal{D}}}_{m(t_{s-1})}(\bm{H}_{m(t_{s-1})})]}{\prod_{j=1}^{m(t_{s-1})}p(A_{j}|\bm{H}_{j})},
\end{eqnarray*}
the Radon-Nikodym derivative ${d \mathcal{P}^{\widetilde{\bm{\mathcal{D}}}_{m(t_{s-1})}}}/{d \mathcal{P}_{m(t_{s-1})}}$ exists under the positivity assumption that $p(A_{j}|\bm{H}_{j}) >0$ for all $1\le j\le m(t_{s-1})$.  Thus, the expectation $E^{\widetilde{\bm{\mathcal{D}}}_{m(t_{s-1})}}$ in \eqref{eq:main} can be expressed as an unconditional one:
\begin{equation}\label{F1}
E^{\widetilde{\bm{\mathcal{D}}}_{m(t_{s-1})}}\left[I(T\ge t_s)\right]= E\left\{ \frac{I[\widetilde{\bm{A}}_{m(t_{s-1})}=\widetilde{\bm{\mathcal{D}}}_{m(t_{s-1})}(\bm{H}_{m(t_{s-1})})]}{p(\tilde{\bm{A}}_{m(t_{s-1})}|\bm{H}_{m(t_{s-1})})} I(T\ge t_s) \right\},
\end{equation}
which incorporates the inverse weighting of the treatment probability $p(\tilde{\bm{A}}_{m(t_{s-1})}|\bm{H}_{m(t_{s-1})})$.
 This probability appears in randomized clinical trials or must to be estimated in observational studies.
Similarly, we have
\begin{equation}\label{F2}
E^{\widetilde{\bm{\mathcal{D}}}_{m(t_{s-1})}}\left[I(T= t_s)\right]= E\left\{ \frac{I[\widetilde{\bm{A}}_{m(t_{s-1})}=\widetilde{\bm{\mathcal{D}}}_{m(t_{s-1})}(\bm{H}_{m(t_{s-1})})]}{p(\tilde{\bm{A}}_{m(t_{s-1})}|\bm{H}_{m(t_{s-1})})} I(T= t_s) \right\}.
\end{equation}
To avoid unobserved confounders for the survival time and treatments, we assume that, given $\bm{H}_{m(t_{s-1})}$, the potential outcomes of $I(T=t_s)$ and $I(T>t_s)$ are independent of $\tilde{\bm{A}}_{m(t_{s-1})}$, which is also a common assumption in inverse probability weighting approaches.

However, the survival time $T$ in (\ref{F1}) and (\ref{F2}) is not fully observed due to censoring. In practice, only the $Y = T \wedge C$ is observed. Note that we have assumed that the censoring time $C$ is independent of $(T, \widetilde{\bm{A}}_m, \widetilde{\bm{X}}_m)$ in Section \ref{sec:2}. In fact, this assumption is standard in clinical trials with several follow-up studies \citep{jiang2017estimation}. Under this non-informative censoring assumption, we can substitute the unobserved survival time $T$ with the observed $Y$ by
\begin{eqnarray*}
	q^{\widetilde{\bm{\mathcal{D}}}_{m_g}}(s)&=&\frac{ E^{\widetilde{\bm{\mathcal{D}}}_{m(t_{s-1})}}\left[I(T=t_s)\right] E[I(C\ge t_s)]}{E^{\widetilde{\bm{\mathcal{D}}}_{m(t_{s-1})}}\left[I(T\ge t_s)\right] E[I(C\ge t_s)]}\\
	&=& \frac{E\{I[\widetilde{\bm{A}}_{m(t_{s-1})} = \widetilde{\bm{\mathcal{D}}}_{m(t_{s-1})}(\bm{H}_{m(t_{s-1})})]I(Y=t_s)\Delta/p(\tilde{\bm{A}}_{m(t_{s-1})}|\bm{H}_{m(t_{s-1})})\}}
	{E\{I[\widetilde{\bm{A}}_{m(t_{s-1})} = \widetilde{\bm{\mathcal{D}}}_{m(t_{s-1})}(\bm{H}_{m(t_{s-1})})]
		I(Y\ge t_s)/p(\tilde{\bm{A}}_{m(t_{s-1})}|\bm{H}_{m(t_{s-1})})\}}.
\end{eqnarray*}
Consequently, we can estimate the survival probability $S^{\widetilde{\bm{\mathcal{D}}}_{m_g}}(t_g)$ with the observed data. That is,
\begin{eqnarray}\label{estimation1}
\bar{S}^{\widetilde{\bm{\mathcal{D}}}_{m_g}}(t_g)
&=& \prod^g_{s=1}\bar{q}^{\widetilde{\bm{\mathcal{D}}}_{m_g}}(s)\notag\\
&=&\prod^g_{s=1} \left\{1-\frac{\sum^N_{i=1} \bar{w}_i(s-1) I(Y_i=t_s)\Delta_i}{\sum^N_{i=1} \bar{w}_i(s-1) I(Y_i\ge t_s)}\right\},
\end{eqnarray}
where
\begin{equation}\label{weight1}
\bar{w}_i(s)=\frac{I[\widetilde{\bm{A}}_{i {m(t_{s})}} = \widetilde{\bm{\mathcal{D}}}_{m(t_{s})}(\bm{H}_{i {m(t_{s})}})]}
{\hat{p}(\tilde{\bm{A}}_{i {m(t_{s})}}|\bm{H}_{i {m(t_{s})}})},
\end{equation}
and $\hat{p}(\tilde{\bm{A}}_{i m(t_{s})}|\bm{H}_{im(t_{s})})$ is an estimator of ${p}(\tilde{\bm{A}}_{i m(t_{s})}|\bm{H}_{i m(t_{s})})$, which depends on the number of possible treatment choices and
can be estimated through a penalized multinomial model in Section \ref{sec:5}.

One crucial component of the proposed method is decomposing the survival function in terms of survival probabilities at time-points before the given $t_g$ as in \eqref{estimation1}. This strategy allows us to only consider treatments received at stages before these time-points. Thus, we can incorporate all available information from patients who are subject to censoring even before the given $t_g$. In addition, our method integrates the estimations of decision rules at multiple stages together instead of recursively fitting the regression model at each stage, which is more robust to model-misspecification.


\subsection{Multicategory dynamic treatment regimes}
  Although many developments have been made in estimation of the optimal DTR for binary treatments, few existing approaches can handle multicategory treatments. In fact, the extension from binary treatment choices to multicategory treatments is nontrivial. For example, one can use a sequential binary classifier such as the one-versus-rest approach, but this could lead to inconsistent estimation \citep{zhang2018multicategory}. In addition, it is difficult to make multicategory treatment comparisons efficiently \citep{tao2017adaptive}.

  To overcome these difficulties, we propose an angle-based weighted Kaplan-Meier method to obtain the optimal DTR $\widetilde{\bm{\mathcal{D}}}_{m_g}^*$, such that $S^{\widetilde{\bm{\mathcal{D}}}_{m_g}^*}(t_g) \ge S^{\widetilde{\bm{\mathcal{D}}}_{m_g}}(t_g)$ for all sequential rules $\widetilde{\bm{\mathcal{D}}}_{m_g}$ under the multicategory treatment framework. We adopt an angle-based classification \citep{zhang2014multicategory} idea to incorporate multicategory treatments.
Under the multicategory {DTR} framework, there are $K$ possible treatments. To visualize the $K$ possible treatments from a geometric perspective, we define a simplex $\bm{V}$ with $K$ vertices $(\bm{V}_1,...\bm{V}_j,...,\bm{V}_K)$ in a $(K-1)$-dimensional space such that
\[
\bm{V}_j=\left\{\begin{array}{ll}
\frac{1}{\sqrt{K-1}}\bm{1}_{K-1}, & j=1,\\
-\frac{1+\sqrt{K}}{(K-1)^{3/2}}\bm{1}_{K-1}  + \sqrt{\frac{K}{K-1}} \bm{e}_{j-1},&2\le j\le K,
\end{array}\right.
\]
where $\bm{1}_{K-1}$ is a vector of $1$'s with a length of $K-1$, and $\bm{e}_j\in \mathbb{R}^{K-1}$ is a vector with $1$ in the $j$-th element and $0$ elsewhere. Here $\bm{V}_j$ represents a potential treatment choice. Since this simplex is symmetric for all vertices, the angle between any pair of vertices $\bm{V}_{j-1}$ and $\bm{V}_j$ is the same.

\begin{figure}
	\centering
	\scalebox{0.3}[0.3]{\includegraphics{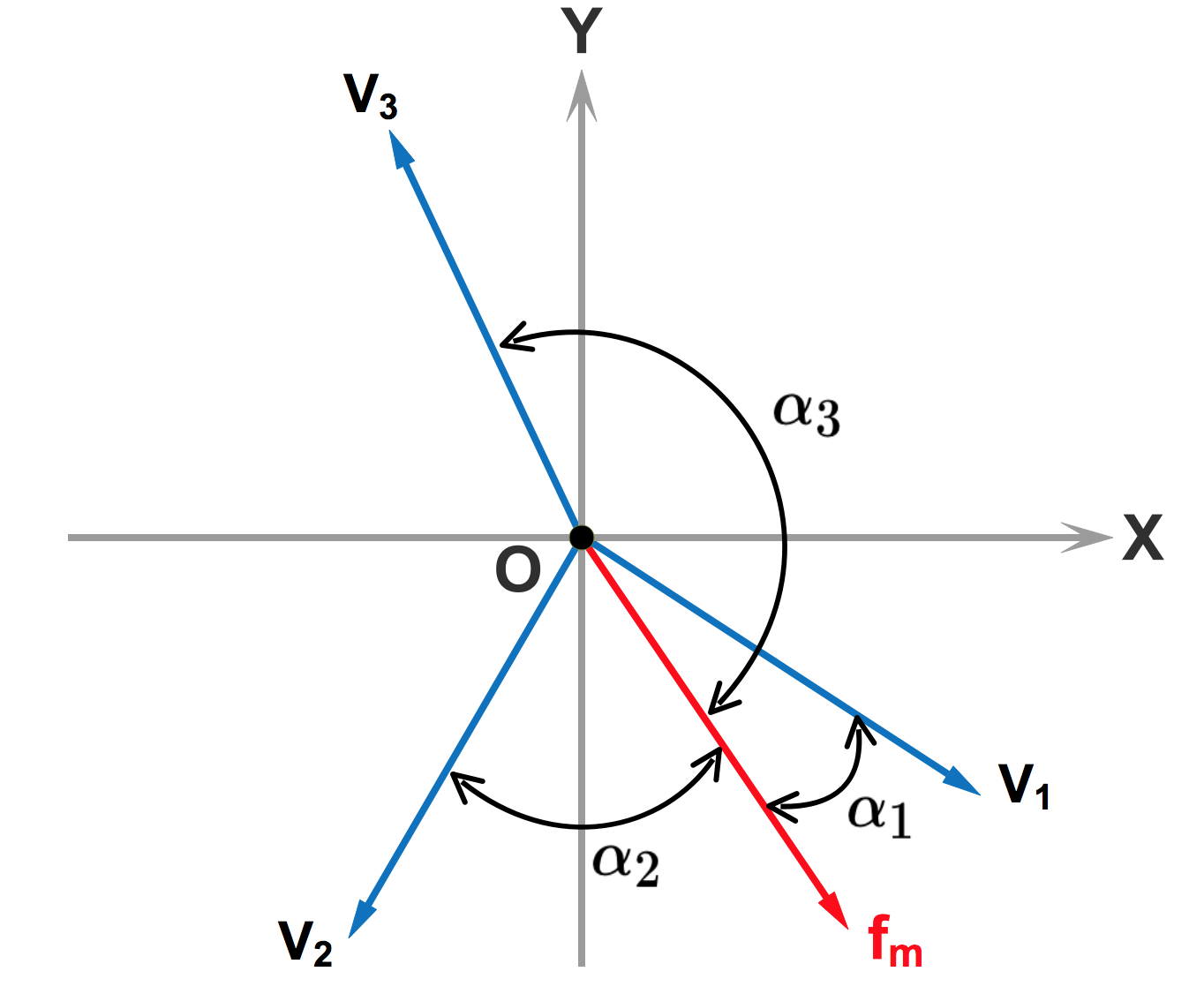}}
	\caption{ Illustration of the angle-based approach with $K=3$. }\label{Angle}
\end{figure}
Under this geometric framework, we utilize the angles between a $(K-1)$-dimensional classification function vector $\bm{f}_{m}(\bm{H}_{m})=\big(f_{{m},1}(\bm{H}_{m} ),\dots, f_{{m}, K-1}(\bm{H}_{m})\big)$ and  $\bm{V}_1,\bm{V}_2,...,\bm{V}_K$ to determine the decision rule at the $m$-th stage for $1\le m \le m_g$.
This decision rule follows a principle which identifies the smallest angle.
Specifically, the decision rule at the $m$-th stage is defined as
\begin{equation}\label{dtr1}
\mathcal{D}_{m}(\bm{H}_{m})=\underset{A\in\{1,\dots,K\}}{\text{argmin}}\text{Ang}(\bm{V}_A,\bm{f}_{m}(\bm{H}_{m}))=\underset{A\in\{1,\dots,K\}}{\text{argmax}}\left<\bm{V}_A,\bm{f}_{m}(\bm{H}_{m})\right>,
\end{equation}
where $\text{Ang}(\bm{V}_A,\bm{f}_{m}(\bm{H}_{m}))$ represents the angle between $\bm{V}_A$ and $\bm{f}_{m}(\bm{H}_{m})$. The last equality in (\ref{dtr1}) follows from the fact that minimizing the angle between $\bm{V}_A$ and $\bm{f}_{m}(\bm{H}_{m})$ is equivalent to maximizing the inner product between $\bm{V}_A$ and $\bm{f}_{m}(\bm{H}_{m})$. 
We let $K=3$ as an illustration example in Figure \ref{Angle}, where $\bm{V}_1$, $\bm{V}_2$, and $\bm{V}_3$ represent the three possible treatment choices. Then the decision corresponding to the classification function $\bm{f}_m$ in Figure \ref{Angle} is treatment $1$, since the angle $\alpha_1$ between $\bm{f}_m$ and $\bm{V}_1$ is the smallest among $\alpha_1$, $\alpha_2$, and $\alpha_3$.

Substituting \eqref{dtr1} for \eqref{weight1}, then maximizing the function in  \eqref{estimation1} becomes a maximization problem of  $\bar{S}^{\widetilde{\bm{\mathcal{D}}}_{m_g}}(t_g)$ over $\bm{f}_{m}$. However, it is still challenging to maximize $\bar{S}^{\widetilde{\bm{\mathcal{D}}}_{m_g}}(t_g)$ directly, since $\bar{S}^{\widetilde{\bm{\mathcal{D}}}_{m_g}}(t_g)$ is discontinuous with respect to $\bm{f}_{m}$ for $1\le m \le m_g$ due to the discontinuity of the indicator function in \eqref{weight1}. Alternatively, a computationally more efficient approach is to replace the indicator by a continuous 
surrogate function. We consider adopting logistic surrogate function
\begin{equation}\label{logistic}
l(u)=\frac{1}{1+e^{-b(u-u_0)}},
\end{equation}
which is a smooth and strictly increasing function,  where {the} coefficient $b > 0$ measures the steepness of the logistic curve and $u_0\in \mathbb{R}$ represents the inflection point.
The choice of the $u_0$ will be discussed in Section \ref{sec:4}.

We propose to replace $I[A_{im}=\mathcal{D}_{im}(\bm{H}_{im})]$ with $l\big( \big<\bm{V}_{A_{im}},\bm{f}_{im}(\bm{H}_{i{m}})\big>\big)$ for $1\le i\le N$  and $1\le m\le m_g$ in \eqref{weight1}. 
That is, the proposed estimator $\hat{\bm{F}}$ is the maximizer of the following penalized log-transformed survival function
\begin{eqnarray}\label{object}
Q_{N,t_g}(\bm{F})&=&\log\left\{\widehat{S}^{\widetilde{\bm{\mathcal{D}}}_{m_g}}(t)\right\} - J_{\lambda}(\bm{F})
 =\log\left\{\prod^g_{s=1}\hat{q}^{\widetilde{\bm{\mathcal{D}}}_{m_g}}(s)\right\} - J_{\lambda}(\bm{F})
 \notag\\
&=&\sum^g_{s=1}\log\left\{1-\frac{\sum^N_{i=1} w_i(s-1) I(Y_i=t_s)\Delta_i}{\sum^N_{i=1} w_i(s-1) I(Y_i\ge t_s)}\right\} 
- J_{\lambda}(\bm{F}),
\end{eqnarray}
where
\begin{equation}\label{stabilized}
w_i(s)= \frac{ \prod_{j=1}^{m(t_s)}  l\left( \left<\bm{V}_{A_{ij}},\bm{f}_{j}(\bm{H}_{j})\right>\right)}{\hat{p}(\widetilde{\bm{A}}_{i{m(t_s)}}|\bm{H}_{i{m(t_s)}})},
\end{equation}
$\bm{F}=(\bm{f}_1, \dots, \bm{f}_{m_g})$, and $J_{\lambda}(\bm{F})$
is a penalty function on $\bm{F}$
with tuning parameter $\lambda$ to prevent overfitting and avoid the identifiability issue.

In Section \ref{sec:5}, we present the implementation of the proposed method,
 and provide the details for tuning the parameters $\lambda$, $b$, $u_0$ and estimation of the propensity score ${\hat{p}(\widetilde{\bm{A}}_{i{m(t_s)}}|\bm{H}_{i{m(t_s)}})}$ in \eqref{stabilized}.

\section{Theory}\label{sec:4}

In this section, we demonstrate the Fisher consistency \citep{liu2007fisher} of the proposed method, indicating that the proposed estimator achieves the optimal DTR if the estimator is calculated using the entire population. 
Although the entire population is typically not observed in practice, nevertheless the Fisher consistency shows that the proposed method is able to predict the best treatment for each subject if we have sufficient information.
Specifically, we show that the DTR corresponding to the proposed estimator $\bar{\bm{F}}$ based on the entire population
recommends the same treatments as the optimal $\widetilde{\bm{\mathcal{D}}}_{m_g}^*$,
where $\bar{\bm{F}}=(\bar{\bm{f}}_1, \dots, \bar{\bm{f}}_{m_g})$ is the maximizer of 
\begin{eqnarray}
Q_{t_g}(\bm{F})=\sum_{s=1}^{g}\log \{\tilde{q}(s, \bm{F})\}-\lambda\sum_{j=1}^{m_g}\sup_{\bm{H}_j}\left\|\bm{f}_j(\bm{H}_j)\right\|_2
\end{eqnarray}
over
$\mathcal{F}=\left\{\bm{F}: \bm{f}_j \text{ is measurable}  \text{ and }
\sup_{\bm{H}_j}\|\bm{f}_j(\bm{H}_j)\|_2<+\infty \; \text{a.s.},
\text{ for } 1\le j \le m_g \right\}$.
Here, we adopt a $L_2$ penalty function and
\begin{eqnarray}\label{popuP}
\tilde{q}(s, \bm{F})=1- \frac{E\left[\prod_{j=1}^{m(t_{s-1})} l\left( \left<\bm{V}_{A_{j}},\bm{f}_j(\bm{H}_{j})\right>\right) I(Y=t_{s})\Delta \big/ \big \{\prod_{j=1}^{m(t_{s-1})}p(A_j|\bm{H}_j)\big\} \right]}{E\left[ \prod_{j=1}^{m(t_{s-1})} l\left( \left<\bm{V}_{A_{j}},\bm{f}_j(\bm{H}_{j})\right>\right) I(Y\ge t_{s})\big/\big\{\prod_{j=1}^{m(t_{s-1})}p(A_j|\bm{H}_j)\big\} \right ]}.
\end{eqnarray}

{\color{black}
	To simplify expression of the following conditions and theorem, we first define some notations. Let $\tilde{\bm{A}}_{w^c}$ be a vector consisting of $A_j$ for $1\le  j \le w-1$ and $w+1\le  j \le m_g$,  and
\begin{eqnarray*}
		&&\hspace{-8mm}\varphi (t_{v(i)}, \tilde{\bm{x}}_{w}, a_w, \tilde{\bm{a}}_{w^c})= E_{w+1}^{\widetilde{\bm{A}}_{w}}\left[\cdots E_{m_g}^{\widetilde{\bm{A}}_{m_g-1}}\left\{P(T>t_{v(i)}|\tilde{\bm{X}}_{w}=\tilde{\bm{x}}_{w}, A_{w}=a_w, \tilde{\bm{A}}_{w^c}=\tilde{\bm{a}}_{w^c})\right\}\cdots\right],\\
		&&\hspace{-8mm}\eta_1(\tilde{\bm{X}}_{w}, \tilde{\bm{A}}_{w^c}, t_{v(i)})= \min_{a_w\ne D^*_w(\bm{H}_w)} \left\{c_L^{i-w} \varphi (t_{v(i)}, \tilde{\bm{X}}_{w}, \mathcal{D}^*_w(\bm{H}_w), \tilde{\bm{A}}_{w^c}) - \varphi (t_{v(i)}, \tilde{\bm{X}}_{w}, a_w, \tilde{\bm{A}}_{w^c})\right\},\\ 
		&&\hspace{-8mm}\eta_2(\tilde{\bm{X}}_{w}, \tilde{\bm{A}}_{w^c}, i)=\max_{a_w\ne D^*_w(\bm{H}_w)} \left\{ \frac{c'_U}{c'_L}  \varphi (t_{v(i)}, \tilde{\bm{X}}_{w}, \mathcal{D}^*_w(\bm{H}_w), \tilde{\bm{A}}_{w^c})
		- c_L^{i-w}  \varphi (t_{v(i)}, \tilde{\bm{X}}_{w}, a_w, \tilde{\bm{A}}_{w^c}) \right\},
	\end{eqnarray*}
where $E_{m}^{\widetilde{\bm{A}}_{m-1}}$ represents an expectation with respect to $\{\bm{X}_m \mid \tilde{\bm{X}}_{m-1}, \widetilde{\bm{A}}_{m-1} \}$ for $w+1\le m \le m_g$; $t_{v(i)}$ is the first time-point at the $i$-th stage for $1\le i\le m_g$, $t_{v(m_g+1)}=t_g$; $c'_L$ and $c'_U$ are the lower bound and upper bound of $l'(\cdot)$, respectively; and $c_L$ is the lower bound of $l(\cdot)$.
\begin{condition}\label{loss}
	The  $l(\cdot)$ in (\ref{popuP}) is a strictly increasing surrogate 
	function such that, $l(0)>\bar{c}$ for some constant $\bar{c}\in (0, 1/2]$, $l(u) \in (0, 1)$ and $l''(u)<0$ for $u\in \left(-|C_Q|/\lambda, |C_Q|/\lambda\right)$, where
	\begin{eqnarray*}
	C_Q=\sum_{s=1}^{g}\log\left\{1- \frac{E\left[\prod_{j=1}^{m(t_{s-1})} I(Y=t_{s})\Delta \big/ \big \{\prod_{j=1}^{m(t_{s-1})}p(A_j|\bm{H}_j)\big\} \right]}{\bar{c}\cdot E\left[ \prod_{j=1}^{m(t_{s-1})} I(Y\ge t_{s})\big/\big\{\prod_{j=1}^{m(t_{s-1})}p(A_j|\bm{H}_j)\big\} \right ]}\right\}.
	\end{eqnarray*}
\end{condition}
\begin{condition}\label{survival_prob}
	For each stage $1\le w \le m_g$, 
	\begin{eqnarray}
	\eta_1(\tilde{\bm{X}}_{w}, \tilde{\bm{A}}_{w^c}, t_g)  - \sum_{i=w+1}^{m_g} \eta_1^-(\tilde{\bm{X}}_{w}, \tilde{\bm{A}}_{w^c},  t_{v(i)})  - \sum_{i=w}^{m_g}\eta_2^+(\tilde{\bm{X}}_{w}, \tilde{\bm{A}}_{w^c}, i)/\kappa_{i} >0,
	\end{eqnarray}
	where
 $\eta^+$ and $\eta^-$ represent the positive part and negative part of $\eta$, respectively, and $\kappa_i$ is the lower bound of $\prod_{j=i+1}^{m_g}p(A_j|\bm{H}_j)$ for $1\le i \le m_g-1$  with  $\kappa_{m_g}=1$.
\end{condition}

Condition \ref{loss} holds when the surrogate function $l(\cdot)$ is positive, strictly increasing, concave, and bounded on a certain area, such as the logistic surrogate function in (\ref{logistic}) with 
	$u_0\le-|C_Q|/\lambda$. Condition \ref{survival_prob} ensures  
that receiving the optimal individualized treatment at each stage increases the targeted survival probability at the given time $t_g$. It also ensures 
that the increase in the survival probability at the given time $t$ exceeds the changes of survival probabilities at the beginning of stage $1\le m \le m_g$.


\begin{theorem}\label{fisher}
	Under Conditions \ref{loss} and \ref{survival_prob}, we have 
	 $$\underset{A\in\{1,\dots,K\}}{\text{argmax}}\left<\bm{V}_A, \bar{\bm{f}}_j(\bm{H}_j)\right>=\mathcal{D}^*_j(\bm{H}_j) \ \ \text{almost surely},$$
	 for stages $j=1, \dots, m_g$.
\end{theorem}

Theorem \ref{fisher} states that the proposed method is Fisher consistent, which is a fundamental property for a classifier. Theorem \ref{fisher} shows that the proposed method recommends the optimal treatment for each subject at each stage with probability $1$ if we obtain sufficient data. To the best of our knowledge, this is the first Fisher consistency result for DTR under the angle-based multicategory framework. The proof of the theorem is provided in the supplementary material.

\section{Implementation}\label{sec:5}

In this section, we demonstrate how to solve the maximization problem with the objective function $Q_{N,t_g}(\bm{\theta})$ in (\ref{object}) and how to select proper tuning parameters.
Specifically, we consider that the classification functions are parametric models $\bm{F}(\bm{H}_{m_g})=\bm{F}(\bm{H}_{m_g},\bm{\theta})$ and we obtain an estimator $\widehat{\bm{\theta}}$ through maximizing $Q_{N,t_g}(\bm{\theta})$ with a $L_2$ penalty, that is,
\begin{equation}\label{problem}
\widehat{\bm{\theta}}=\arg\max_{\bm{\theta}}Q_{N,t_g}(\bm{\theta}),
\end{equation}
where $Q_{N,t_g}(\bm{\theta})=Q_{N,t_g}(\bm{F}(\bm{H}_{m_g},\bm{\theta}))$ and the penalty part of $Q_{N,t_g}(\bm{\theta})$ is $J_{\lambda}(\bm{F}(\bm{H}_{m_g},\bm{\theta}))=\|\bm{\theta}\|_2$.
Here, we use the gradient descent algorithm to solve \eqref{problem}, and the explicit algorithm can be summarized in the following table:

\noindent $\overline{\mbox{\underline{\makebox[\textwidth]{\textbf{Implementation:} Gradient descent algorithm}}}}$
\begin{enumerate}
	\item[1.] (Initialization) Input all observed $\{(Y_i,\Delta_i,\bX_{im_j},A_{im_j}): m_j=1,\ldots,m_g\}^N_{i=1}$, the initial step size $\eta_0$, initial value $\bm{\theta}_0$, and a stopping criterion $\varepsilon=10^{-4}$.
	
	\item[2.] (Iteration) At the $(k+1)$-th iteration $(k\ge0)$, estimate $\bm{\theta}_{k+1}$ as follows
	$$
	\widehat{\bm{\theta}}_{k+1} = \widehat{\bm{\theta}}_k + \eta_k\nabla Q_{N,t_g}(\widehat{\bm{\theta}}_k),
	$$
	where $\nabla Q_{N,t_g}(\widehat{\bm{\theta}}_k)$ is the gradient value of $Q_{N,t_g}(\bm{\theta})$ at $\widehat{\bm{\theta}}_k$, and the step size $\eta_k$ is chosen by the Barzilai--Borwein method \citep{barzilai1988two}; that is,
	$$
	\eta_k=\frac{(\widehat{\bm{\theta}}_k-\widehat{\bm{\theta}}_{k-1})^{\T}\{\nabla Q_{N,t_g}(\widehat{\bm{\theta}}_k)
		-\nabla Q_{N,t_g}(\widehat{\bm{\theta}}_{k-1})\}}
	{\|\nabla Q_{N,t_g}(\widehat{\bm{\theta}}_k)
		-\nabla Q_{N,t_g}(\widehat{\bm{\theta}}_{k-1})\|^2}.
	$$

	\item[3.] (Stopping Criterion) Stop if
	$\max\{\|\widehat{\bm{\theta}}_{k+1}-\widehat{\bm{\theta}}_k\|^2,|\nabla Q_{N,t_g}(\widehat{\bm{\theta}}_{k+1})
	-\nabla Q_{N,t_g}(\widehat{\bm{\theta}}_{k})|^2\}<\varepsilon$.
	Otherwise go to step 2.
	
\end{enumerate}
\noindent\makebox[\linewidth]{\rule{\textwidth}{0.4pt}}

In practice, a proper choice of the parameters $u_0$, $b$ in (\ref{logistic}) and 
$\lambda$ in (\ref{object}) could enable our algorithm to estimate the optimal {DTR}. In particular, the choice of the tuning parameter $\lambda$ controls model complexity. We let $u_0=-|C_Q|/\lambda$ to satisfy Condition \ref{loss} for any given $b$ and $\lambda$. Various tuning techniques have been put forward in the existing literature. Here we briefly discuss the $d$-fold cross-validation procedure, which is commonly used in practice. Specifically, all the samples are randomly partitioned into $d$ equally sized subsets.
For each pair of tuning parameters $(b, \lambda)$, we fit the model based on every $d-1$ subsets, and calculate the estimated survival function in (\ref{estimation1}) at the given time $t_g$ for the remaining subset. Equivalently, the $d-1$ subsets form a training set, while the remaining subset serves as a validation set. We obtain a total of $d$ training sets and $d$ validation sets, and choose the optimal tuning parameters which maximize the average of the estimated survival probabilities over all of the validation sets. That is,
$$
(\hat{b},\hat{\lambda})_{\rm validated}=\underset{{(b,\lambda)}}{\text{argmax}}\frac{1}{d}\sum_{r=1}^d Q^r_{N, t_g}(\widehat{\bm{\theta}}|b,\lambda),
$$
where $\widehat{\bm{\theta}}$ is the proposed estimator given $b$ and $\lambda$ using the $r$-th training set, and
$Q^r_{N, t_g}(\widehat{\bm{\theta}}|b,\lambda)$ is the estimated survival probability at the given time $t_g$ based on $\hat{\bm{\theta}}$ and the corresponding validation set.

For data from an observational study, at the $m$-th stage ($1\le m \le m_g$), the propensity scores $P(A_{im}|\bm H_{im})$ in \eqref{weight1} and \eqref{stabilized} are unknown and thus need to be estimated. In this paper, we use a penalized multinomial model to estimate the propensity scores.
At the {$m$-th} stage, we define $\beta_{0 k}$ as the $k$-th intercept for $k=1,...,K$, and $\bm{\beta}_{p_m \times K}$ as the $p_m \times K$ coefficient matrix, where $p_m$ is the dimension of $\bm H_m$. Let $\bm{\beta}_k$ be the $k$-th column of $\bm{\beta}_{p_m \times K}.$ Then we model
$$
P(A_m =k | \bm H_m)=\frac{\exp ({\beta_{0 k}+\bm{\beta}_{k}^{\T} \bm H_m})}{\sum_{j=1}^{K} \exp ({\beta_{0 j}+\bm{\beta}_{j}^{\T} \bm H_m})}.
$$
The log-likelihood function of this model is
\begin{eqnarray*}
\log \big[L(\big \{\beta_{0 k},\bm{\beta}_{k}\}_{1}^{K}\big)\big] = \frac{1}{N} \sum_{i=1}^{N}\left\{\sum_{k=1}^{K} I \big(A_{im}=k \big)\left(\beta_{0 k}+ \bm H_{im}^{\T} \bm{\beta}_{k}\right)-\log \left(\sum_{k=1}^{K} e^{({\beta_{0k}+ \bm H_{im}^{\T} \bm{\beta}_{k}})}\right)\right\}.
\end{eqnarray*}
To obtain estimators for $\beta_{0k}$ and $\bm{\beta}_k$ ($1\le k \le K$), we minimize the penalized negative log-likelihood:
$$
\underset{\{\beta_{0 k}, \bm{\beta}_{k}\}_{1}^{K}}{ \min} -\log \Big[L\big( \{\beta_{0 k}, \bm{\beta}_{k}\}_{1}^{K}\big)\Big] +\lambda^{*} \sum_{k=1}^{K}\left\|\bm{\beta}_{k}\right\|_1,
$$
where $\|\cdot \|_1$ is the $L_1$ norm, and $\lambda^{*}$ is the tuning parameter to control the overall strength of the penalty and can be tuned by cross-validation.

\section{Simulation Studies}\label{sec:6}
In this section, we construct four simulation examples with both linear and non-linear classification functions to assess the finite sample performance of the proposed method. For all examples,
we compare the proposed method with the censored Q-learning method \citep{goldberg2012q} and the subgroup identification method (subgroup) \citep{huling2018subgroup}. Since the censored Q-learning and subgroup methods can only handle binary treatments, we extend these
methods through the one-versus-rest approach to handle multicategory treatments
 for fair comparisons. Specifically, we conduct sequential 0-1 binary treatment estimations, i.e., the treatment $k$ versus others $(k=1,...,K)$, using the censored Q-learning and subgroup methods. If the $k$-th binary classifier recommends $1$, then the treatment $k$ will be selected as the recommended treatment. For the subgroup method, since it is only designed to estimate the optimal decision rule at a single stage, we use it to recommend the optimal treatment at each single stage separately, and then combine the recommended treatments together as the recommended sequential treatments. We assess the performance of these methods via the estimated conditional survival probability under each estimated DTR as defined in \eqref{estimation1}. 

We generate data $(Y_i,C_i,T_i,\Delta_i,\bm X_{im},A_{im})$ for $i=1,\ldots,N$ and $m=1,\ldots,M$, where $M$ is the number of total stages, $\bm X_{im}$ consists of $p$ covariates for the $i$-th patient at the $m$-th stage, and $A_{im}$ is the treatment assigned to the $i$-th patient at the $m$-th stage. For all of our simulation examples, we consider $p=25$, $M=5$, and {three potential treatment choices \{$1, 2, 3$\}}. We first generate a training set with the sample size of $500$ for the model fitting, and then we simulate an independent testing set with $2000$ observations to evaluate the model performance. We consider three different censoring rates, and specify two target time-points $t_g=1.4$ and $2.1$ which correspond to the 3rd and $5$-th stages, respectively.
We perform simulations 150 times and report the average values of estimated conditional survival probabilities. We also provide boxplots under the three different censoring rates for evaluation of the numerical performance. The details of each setting are described as follows:

$\textbf{Example 1}$: {In this example, we let the underlying sequential decision rule $d_{im}$ and the true assigned treatment $A_{im}$ both follow an independent discrete $\text{Uniform}\{1,2,3\}$ distribution for $1\le i \le n$ and $1\le m \le 5$. We also let $\bm{c}_{1m} = (0.5m,0,0,...,0), \bm{c}_{2m} = (0,0.5m,0,...,0)$, and $\bm{c}_{3m} = (0,0,0.5m,0,...,0)$ be three $p$-dimensional vectors. We generate covariates $\bm X_{im}$ from a multivariate normal distribution $\mathcal{N}(\bm{c}_{jm}, 0.1\bm{I}_{p})$, where $\bm{I}_p$ is a $p$ dimensional identity matrix, and $j=d_{im}$.}
The true survival time for each subject is 
$$T_i = \exp\left\{\sum_{m=1}^M0.5I(A_{im}=d_{im})-3X_{i15}+X^3_{i22} - |X_{i33}|+\epsilon_{1i}\right\},$$
and the censoring time is $C_i = \exp\left(0.5\epsilon^2_{2i}-\epsilon_{2i}+ \epsilon_{3i}-2\epsilon^2_{4i}+c_0\right)$,
where $\epsilon_{ri}$ ($r=1,2,3,4$) are generated from an independent normal distribution $\mathcal{N}(0,1)$, and $c_0$ is a constant added to control the censoring rate.

Example 1 is designated to evaluate the numerical performance of the proposed method when the decision rules are relatively simple and have immediate treatment effects. Note that although the decision rules $d_{im}$ are uniformly generated, the covariates $\bm X_{im}$ are related to the decision rule since the mean of $\bm X_{im}$ depends on  $d_{im}$. According to Figures \ref{Fig:stage2test3} and \ref{Fig:stage2test5}, the proposed method performs much better than the competing methods at both stages 3 and 5. Also, Table \ref{stage2table} shows that under the $74\%$ censoring rate and $t_g=2.1$, the proposed method yields $35.3\%$ and $16.6\%$ improvement compared to the censored Q-learning and subgroup methods, {respectively, in terms of the conditional survival probabilities. This} implies that the proposed method is a better performer when the underlying sequential decision rules have a relatively simple form.

\begin{table}  [H]
	\begin{center}
		\caption{{{Example 1}: The estimated conditional survival probability $\bar{S}^{\widehat{\bm{\mathcal{D}}}}(t_g)$ with standard deviations (SD) in parentheses based on 150 simulations. ``Imp-rate'' represents the improvement rate of the proposed method compared to other methods.}} \label{stage2table}

		\renewcommand{\arraystretch}{1.2} \tabcolsep 0.1in 
		\begin{tabular}{cccccccccccccccccc}  \hline  \hline
			Stage & $t_g$ & Censoring Rate  &Method & $\bar{S}^{\widehat{\bm{\mathcal{D}}}}(t_g)$(SD) & Imp-rate \\ \hline
3 & 1.4 & 74\% & Proposed	   &   \textbf{0.674}(0.072)  & -- \\
                                                   &   &   &Qlearning &  0.539(0.080) & 24.9\% \\
		        &    &  &Subgroup &   0.591(0.087) & 14.0\%  \\

			5 & 2.1 & 74\% & Proposed	   &  \textbf{0.586}(0.142)  & --\\
			                      &  &    &Qlearning & 0.433(0.186) & 35.3\%  \\
			        &  &    &Subgroup & 0.503(0.153) &  16.6\%  \\
			\hline
		\end{tabular}
	\end{center}
\end{table}

\begin{figure} [H]
   \begin{minipage}{0.4\textwidth}
     \centering
     \includegraphics[width=1\linewidth]{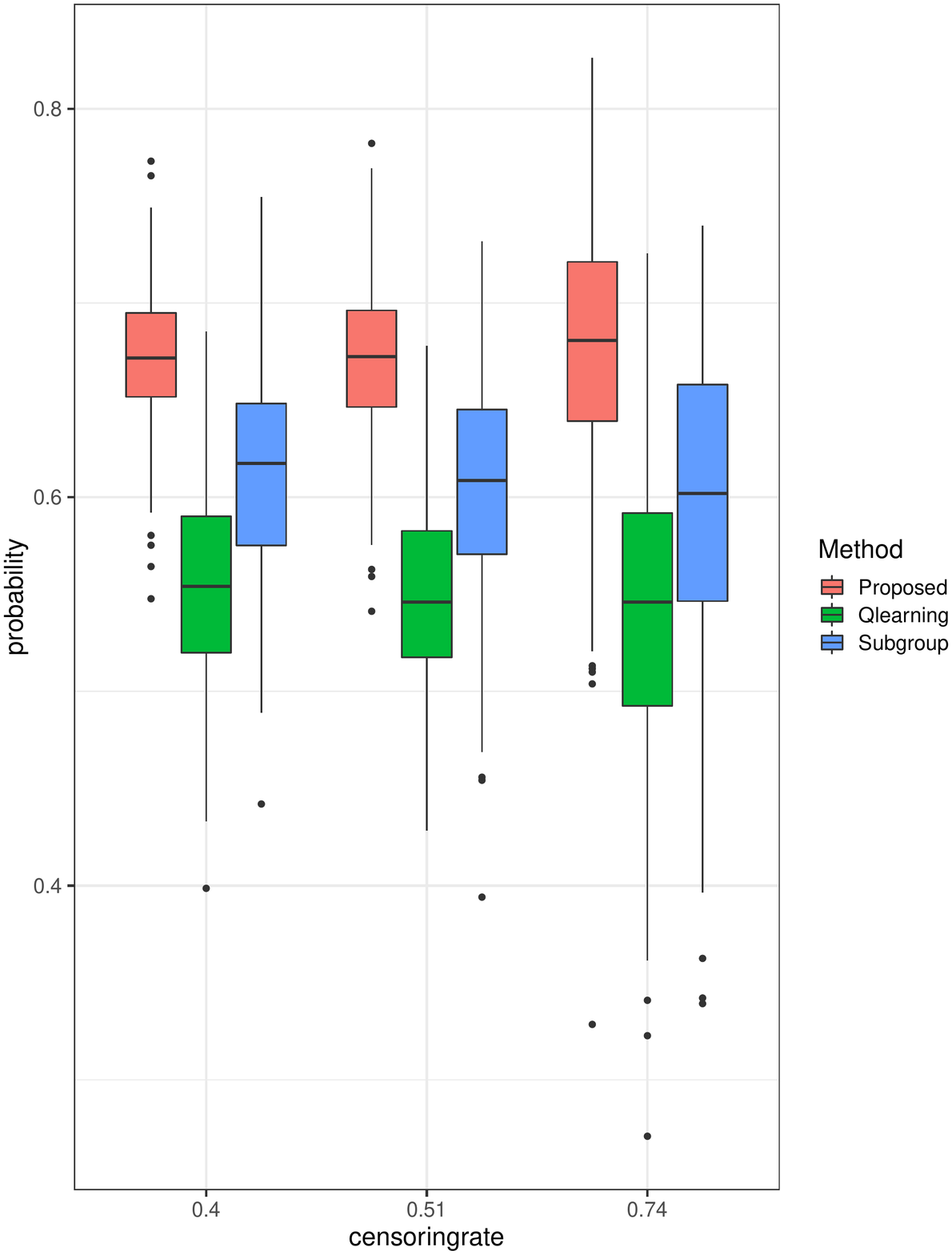}
         \caption{{Example 1}: {Boxplots of the estimated conditional survival probabilities at the third stage for different censoring rates.}}\label{Fig:stage2test3}
   \end{minipage}\hfill
   \begin{minipage}{0.4\textwidth}
     \centering
     \includegraphics[width=1\linewidth]{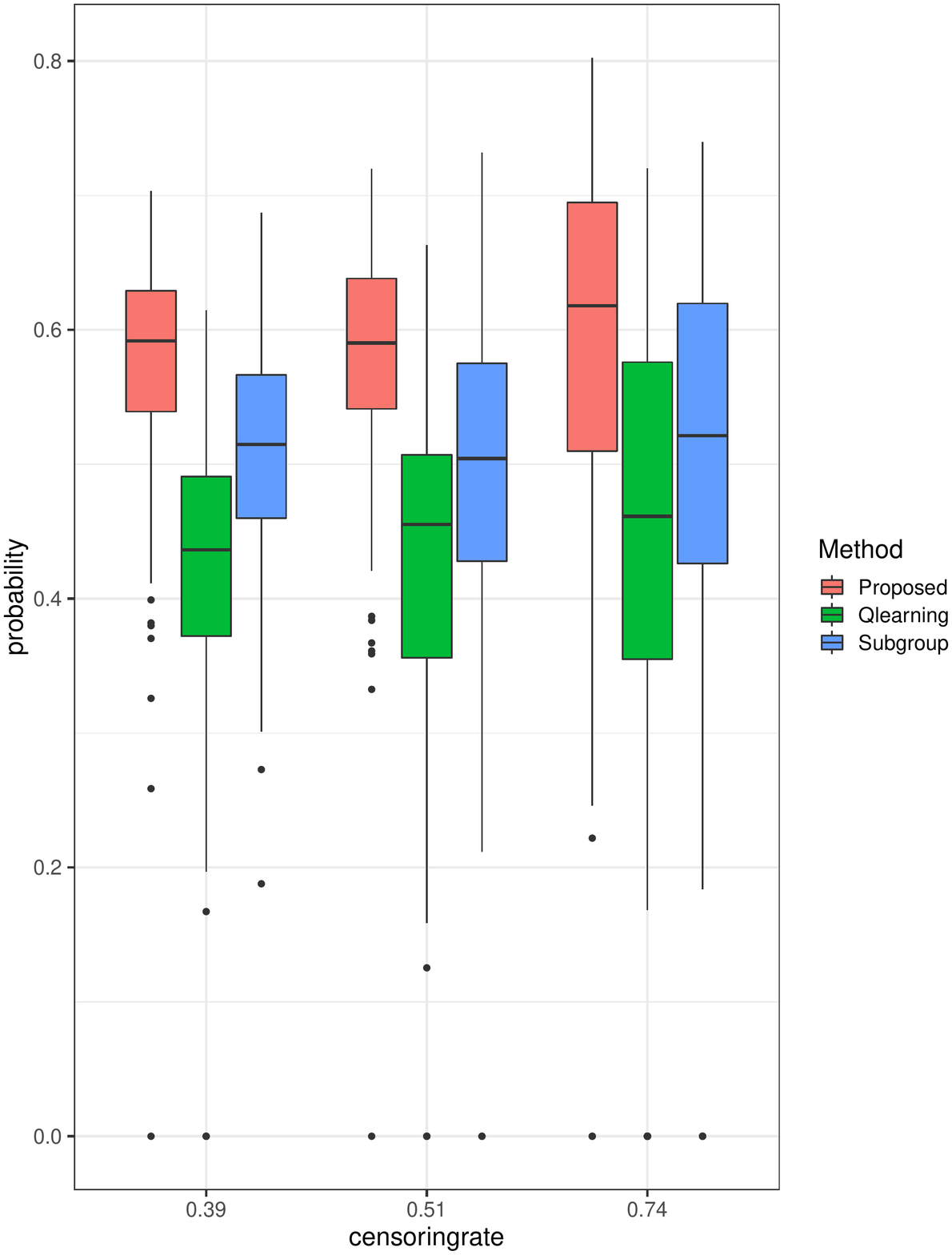}
     \caption{ {Example 1}: {Boxplots of the  estimated conditional survival probabilities at the fifth stage for different censoring rates.}}\label{Fig:stage2test5}
   \end{minipage}
\end{figure}

$\textbf{Example 2}$: {For the $i$-th subject at the $m$-th stage, the decision rule $d_{im}$ is determined by the following classification function vector 
$$\bm{f_m}(\bm H_{im}) = \left(f_{1m}(\bm H_{im}),f_{2m}(\bm H_{im})\right)^{\T}=\left(\bm H_{im}^{\T}{\bm{\theta}_{m1}}, \bm H_{im}^{\T}{\bm{\theta}_{m2}}\right)^{\T}.$$
The $d_{im}$ is defined as
$$
d_{im}= \underset{{{k \in\{1,2,3\}}}}{\text{argmin}}\; {\rm Ang} \{\bm{f_m}(\bm H_{im}),\bm{V}_k\}.
$$
Let $\bm{\theta}_{1j} = {\bm{\theta}_j}^{\T}$,  $\bm{\theta}_{2j} =  (0.5\bm{\theta}_j, -1.5\bm{\theta}_j, 0.1)^{\T}$, $\bm{\theta}_{3j} = (0.25\bm{\theta}_j, -0.5\bm{\theta}_j, \bm{\theta}_j, 0.05, 0.1)^{\T}$, $\bm{\theta}_{4j} = (0.1\bm{\theta}_j, -0.25\bm{\theta}_j, 0.5\bm{\theta}_j, \bm{\theta}_j, 0.01, -0.05, 0.1)^{\T}$, and $\bm{\theta}_{5j} = (0.05\bm{\theta}_j, -0.1\bm{\theta}_j, 0.25\bm{\theta}_j, 0.5\bm{\theta}_j, \bm{\theta}_j,  0.01, \linebreak -0.05, 0.05, -0.1)^{\T}$ for $j=1,2$, where $\bm{\theta}_1 = (1,1,1,0,...,0)$ and $\bm{\theta}_2 = (1,-1,-1,0,...,0)$ are both $p$-dimensional vectors.} We generate covariates $\bm X_{im}$ from a multivariate normal distribution $\mathcal{N}(\bm{0},\bm{I}_p)$. The assigned treatment $A_{im}$ follows an independent discrete $\text{Uniform}\{1,2,3\}$ distribution. The true survival time for the $i$-th subject is
$$T_i =  \exp\left\{\sum_{m=1}^M0.75I(A_{im}=d_{im})-0.5|X_{i15}|+X_{i12}+\epsilon_{1i}\right\},$$
 and the censoring time is $C_i = \exp\left(0.5\epsilon_{2i} + \epsilon_{3i}-\epsilon_{4i} +c_0\right)$, where $\epsilon_{ri}$ ($r=1,2,3,4$) are generated from the independent normal distribution $\mathcal{N}(0,1)$, and $c_0$ is a constant added to control the censoring rate.

Example 2 represents a case with linear classification functions and angle-based decision rules, which is more complex than Example 1 where decision rules are randomly generated.
	In this situation, the proposed method can still outperform the competing methods. According to Table \ref{stage1table}, 
when $t_g=2.1$ with $66\%$ censoring rate, the improvement rates of the proposed method are $33.9\%$ and $29.1\%$ compared to the censored Q-learning and the subgroup methods, respectively. In contrast to the subgroup method, which only incorporates patients' information from a single stage to estimate the optimal decision rule for that stage, the propose method utilizes information from all sages, and can thus achieve better performance. In addition, the improvement of the proposed method compared to the censored Q-learning method is likely because the proposed method can directly estimate all decision rules simultaneously. {Yet} the censored Q-learning method
recursively models the relationship between the survival time and covariates for measuring treatment effects, which requires an additional step of fitting the model and
could cause bias under the model misspecification.

\begin{table}  [H]
	\begin{center}
		\caption{{{Example 2}: The estimated conditional survival probability $\bar{S}^{\widehat{\bm{\mathcal{D}}}}(t_g)$ with standard deviations (SD) in parentheses based on 150 simulations. ``Imp-rate'' represents the improvement rate of the proposed method compared to other methods.}} \label{stage1table}

		\renewcommand{\arraystretch}{1.2} \tabcolsep 0.1in 
		\begin{tabular}{cccccccccccccccccc}  \hline  \hline
			Stage & $t_g$ & Censoring Rate  &Method & $\bar{S}^{\widehat{\bm{\mathcal{D}}}}(t_g)$(SD)  & Imp-rate \\ \hline
3 & 1.4 & 61\% & Proposed	   &   \textbf{0.857}(0.071)  & --\\
                                                   &   &   &Qlearning & 0.660(0.057) & 29.8\% \\
		        &    &  &Subgroup &  0.687(0.069)& 24.7\% \\

5 & 2.1 & 66\% & Proposed	   &   \textbf{0.731}(0.097)  & -- \\
                                                   &   &   &Qlearning & 0.546(0.108) & 33.9\% \\
		        &    &  &Subgroup &  0.566(0.137) & 29.1\% \\
			\hline
		\end{tabular}
	\end{center}
\end{table}

\begin{figure} [H]

   \begin{minipage}{0.4\textwidth}
     \centering
     \includegraphics[width=1\linewidth]{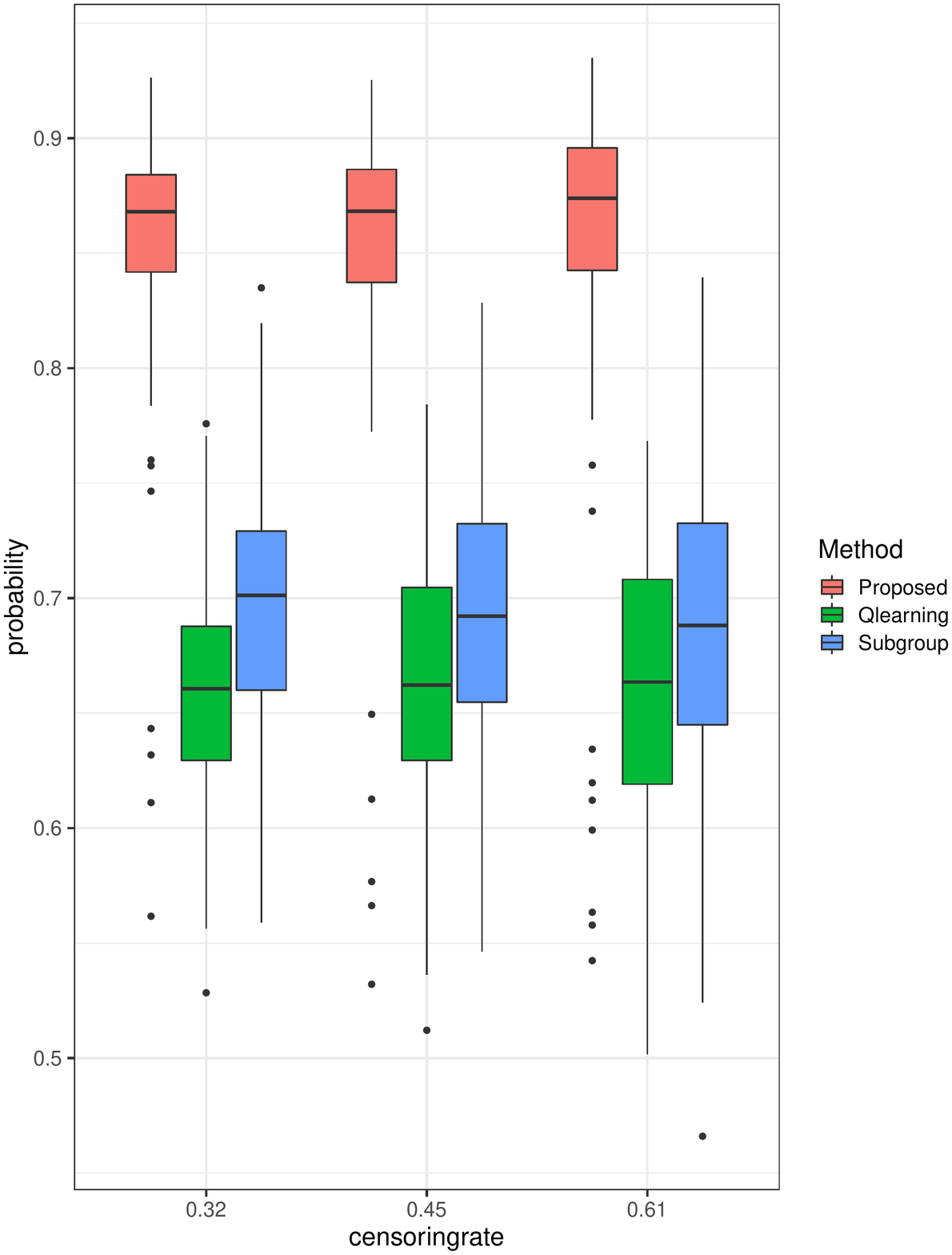}
     \caption{ {Example 2}: {{Boxplots of the  estimated conditional survival probabilities at the third stage for different censoring rates.}}}\label{Fig:stage1test3}
   \end{minipage}\hfill
   \begin{minipage}{0.4\textwidth}
     \centering
     \includegraphics[width=1\linewidth]{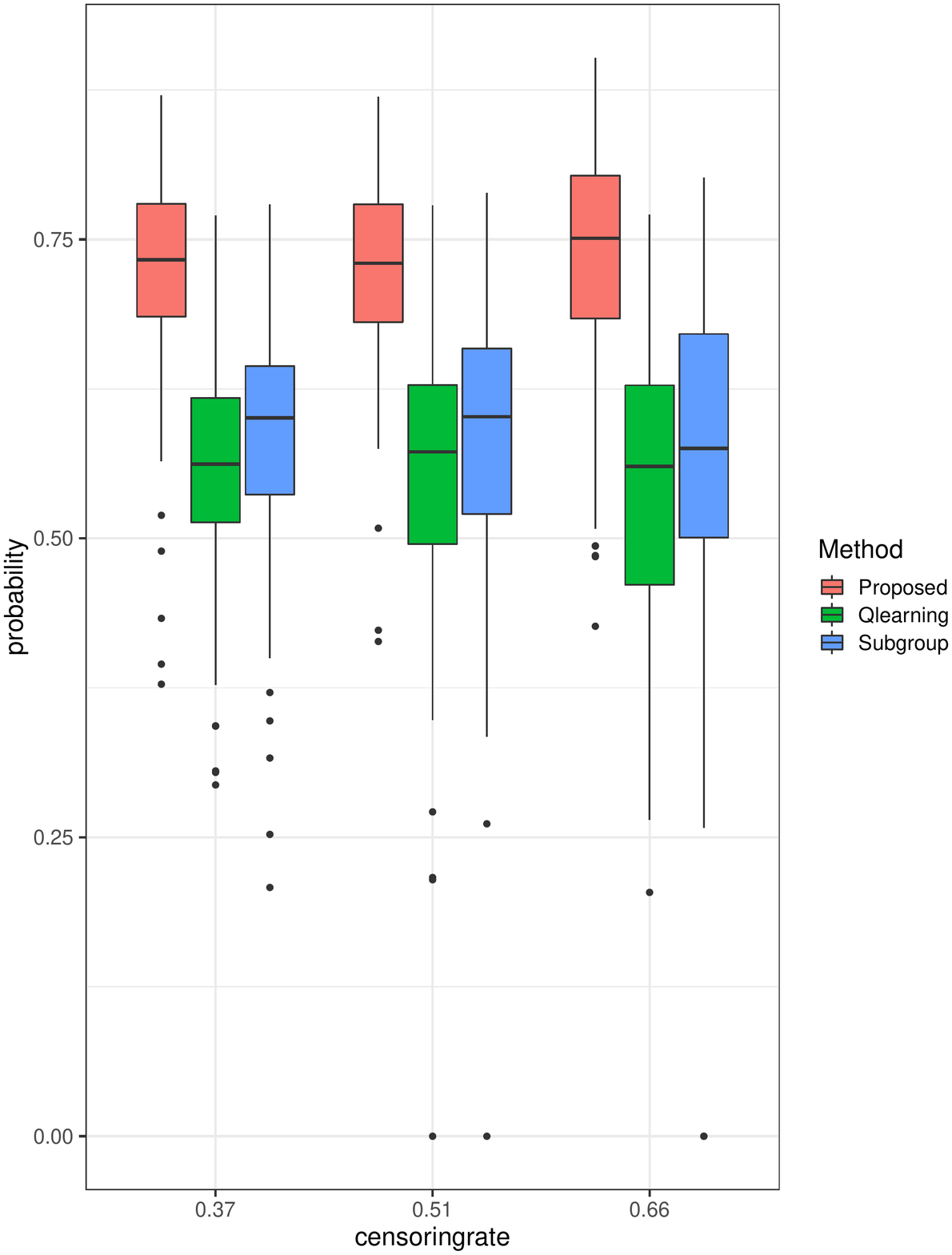}
     \caption{ {Example 2}: {{Boxplots of the  estimated conditional survival probabilities at the fifth stage for different censoring rates.}}}\label{Fig:stage1test5}
   \end{minipage}
\end{figure}

{$\textbf{Example 3}$: We follow the same setting as the one in Example 2, except that we do not make the distribution of $A_{im}$ follow a uniform distribution.
	In particular, we specify the true propensity score as follows:
 \begin{eqnarray*}
 P(A_{im} = 3| {\bm H}_{im}) &=&\Bigg(1+\sum^{2}_{j=1} \exp({{\bm H}_{im}}^{\T}{\bm{\gamma}_{jm}})\Bigg)^{-1} \\
\text{and} \qquad P(A_{im} = k|{\bm H}_{im}) &=& \frac{\exp({{\bm H}_{im}}^{\T}{\bm{\gamma}^{}_{km}})}{1+\sum^{2}_{j=1} \exp({{\bm H}_{im}}^{\T}{\bm{\gamma}_{jm}})},
 \end{eqnarray*}
where $k=1,2$, $m=1,....5$, $\bm{\gamma}_{jm}= (\bm{\gamma}_j,...,\bm{\gamma}_j,\bm{0}_{m-1})^{\T}$, which includes $m$ copies of $p$-dimensional vector $\bm{\gamma}_j$ with $\bm{\gamma}_1 = (0,0,0.25,0,...,0)$ and $\bm{\gamma}_2 = (0,-0.25,0.5,0,...,0)$, and $(m-1)$-dimensional zero vector $\bm{0}_{m-1}$.

In Example 3, to mimic observational studies, the treatments are not uniformly assigned and the probabilities corresponding to treatment assignments (the propensity scores) are unknown. 
We utilize the penalized multinomial model proposed in Section \ref{sec:5} to estimate the propensity scores.
Table \ref{stage3table} and Figure \ref{Fig:stage3test3}-\ref{Fig:stage3test5} indicate that the proposed method still leads to larger estimated conditional survival probabilities than the two existing methods. The improvement rates compared to the censored Q-learning and subgroup method are about $28.5\%$ and $23.1\%$, respectively, for $t_g=1.4$.

\begin{table}  [H]
	\begin{center}
		\caption{{{Example 3}: The estimated conditional survival probability $\bar{S}^{\widehat{\bm{\mathcal{D}}}}(t_g)$ with standard deviations (SD) in parentheses based on 150 simulations. ``Imp-rate'' represents the improvement rate of the proposed method compared to other methods.}} \label{stage3table}
		\renewcommand{\arraystretch}{1.2} \tabcolsep 0.1in 
		\begin{tabular}{cccccccccccccccccc}  \hline  \hline
			Stage & $t_g$ & Censoring Rate  &Method & $\bar{S}^{\widehat{\bm{\mathcal{D}}}}(t_g)$(SD)  & Imp-rate \\ \hline
3 & 1.4 & 61\% & Proposed	   &   \textbf{0.846}(0.078)   & --\\
                                                   &   &   &Qlearning & 0.659(0.057)  & 28.5\% \\
		        &    &  &Subgroup &  0.688(0.063) & 23.1\% \\

5 & 2.1 & 66\% & Proposed	   &   \textbf{0.728}(0.122)  & -- \\
                                                   &   &   &Qlearning & 0.570(0.116)  & 27.6\% \\
		        &    &  &Subgroup &  0.567(0.138) & 28.3\% \\
			\hline
		\end{tabular}
	\end{center}
\end{table}

%

\begin{figure} [H]

   \begin{minipage}{0.4\textwidth}
     \centering
     \includegraphics[width=1\linewidth]{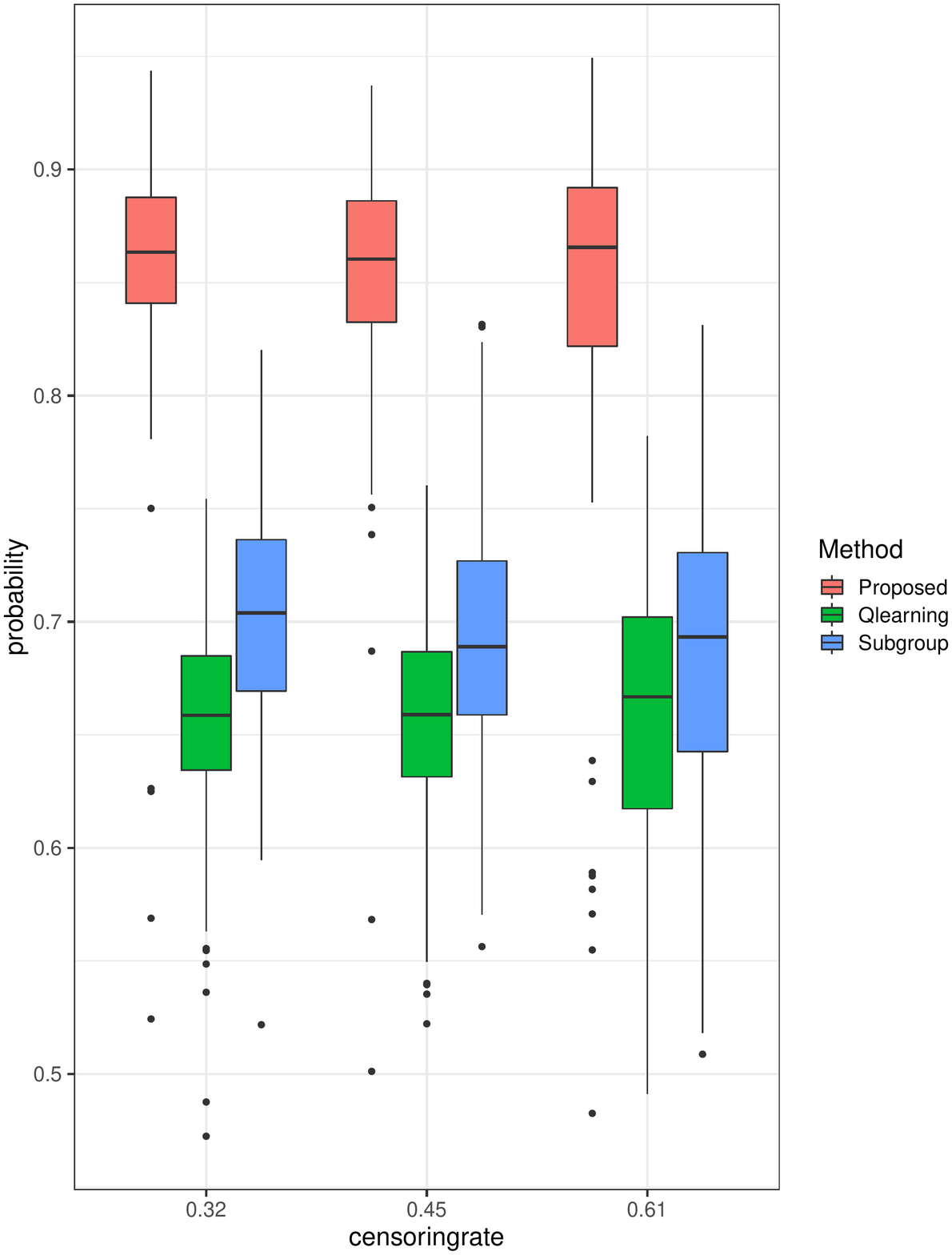}
     \caption{{Example 3}: {Boxplots of the  estimated conditional survival probabilities at the third stage for different censoring rates.}}\label{Fig:stage3test3}
   \end{minipage}\hfill
   \begin{minipage}{0.4\textwidth}
     \centering
     \includegraphics[width=1\linewidth]{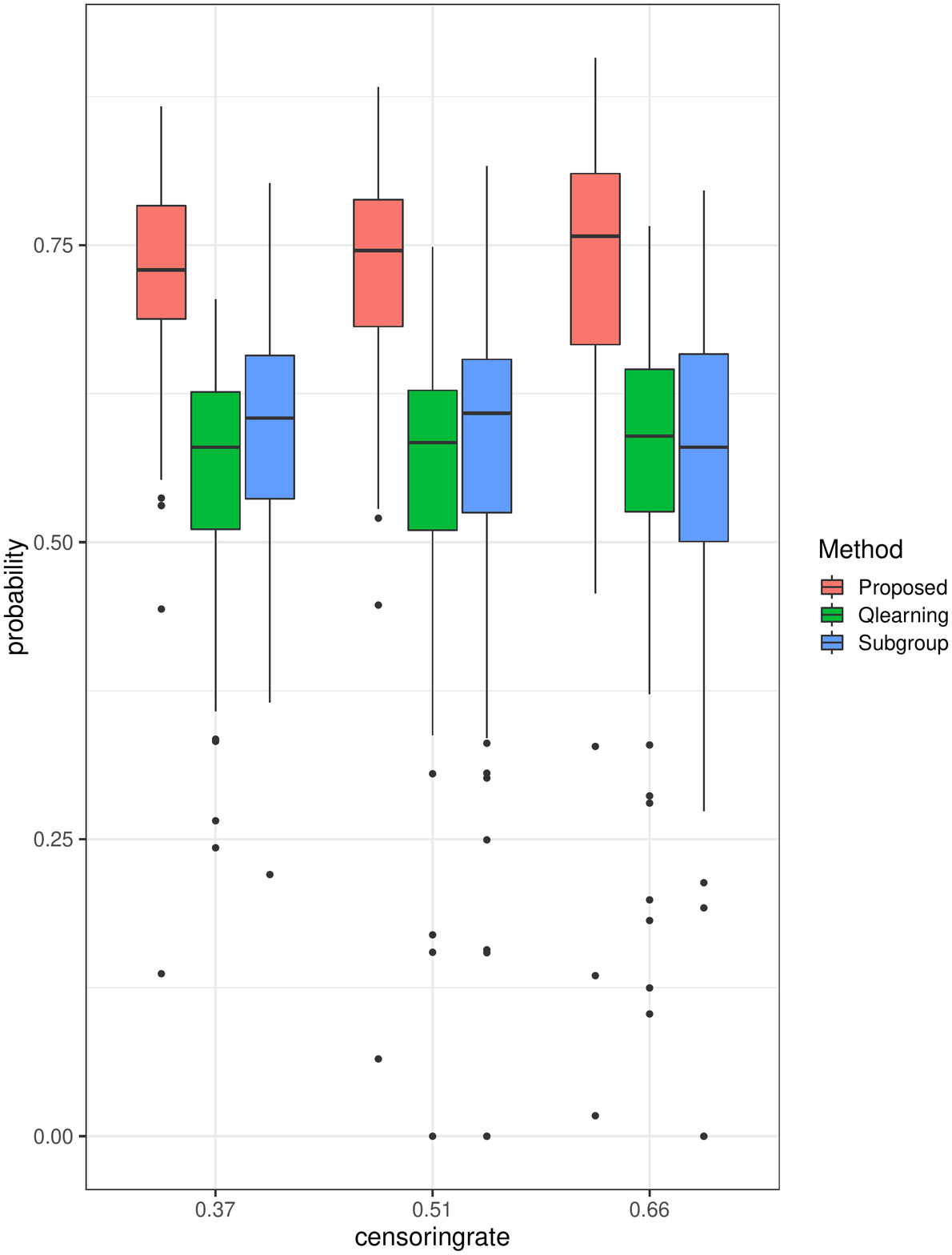}
     \caption{{Example 3}: {Boxplots of the estimated conditional survival probabilities at the fifth stage for different censoring rates.}}\label{Fig:stage3test5}
   \end{minipage}
\end{figure}

$\textbf{Example 4}$: {In this example, we let $f_{1m}(\bm H_{im}) = (\bm H_{im}^{\T}{\bm{\theta}_{m1}})^3$ \text{and} $f_{2m}(\bm H_{im}) = (\bm H_{im}^{\T}{\bm{\theta}_{m2}})^3$ for $1\le i \le n$ and $1\le m \le 5$.
Also, we let  $\bm{\theta}_{1j} = {\bm{\theta}_j}^{\T}$, $\bm{\theta}_{2j} =  (-0.5\bm{\theta}_j, \bm{\theta}_j, -0.3)^{\T}$, $\bm{\theta}_{3j} = (0.25\bm{\theta}_j, -0.25\bm{\theta}_j, 1.5\bm{\theta}_j, -0.05, -0.1)^{\T}$, $\bm{\theta}_{4j} = (0.25\bm{\theta}_j, -0.25\bm{\theta}_j, -0.25\bm{\theta}_j, \bm{\theta}_j, 0.05, 0.05, -0.15)^{\T}$, and $\bm{\theta}_{5j} = (0.05\bm{\theta}_j, -0.1\bm{\theta}_j, -0.25\bm{\theta}_j, -0.5\bm{\theta}_j, \bm{\theta}_j, 0.05, -0.05, 0.05, -0.1)^{\T}$ for $j=1,2$, where $\bm{\theta}_1 = (-1,0.5,-1,0,...,0)$ and $\bm{\theta}_2 = (0.5,-1,-1,0,...,0)$ are $p$-dimensional vectors. The
decision rule $d_{im}$ is defined as
\begin{equation}
d_{im}=\left\{
\begin{aligned}
\notag
1+\big[\text{sgn}\{f_{1m}(\bm H_{im})\}\big]^{+} &+\big[\text{sgn}\{f_{2m}(\bm H_{im})\}\big]^{+}, \qquad \; \; \text{with probability 0.95}, \\
&U_i,  \qquad \qquad \qquad \qquad \qquad \quad    \text{with probability 0.05},
\end{aligned}
\right.
\end{equation}
where $U_i$ follows a discrete $\text{Uniform}\{1,2,3\}$ distribution independent of $(A_{im},\bm X_{im})$, and 
 $\big[\text{sgn}\{f_{1m}(\bm H_{im})\}\big]^{+}$ and $\big[\text{sgn}\{f_{2m}(\bm H_{im})\}\big]^{+}$ represent the positive parts of $\text{sgn}\{f_{1m}(\bm H_{im})\}$ and $\text{sgn}\{f_{2m}(\bm H_{im})\}$, respectively. Each covariate in} $\bm X_{im}$ follows a continuous uniform distribution $U(0,1)$. The assigned treatment $A_{im}$ follows an independent discrete $\text{Uniform}\{1,2,3\}$ distribution. The true survival time for the $i$-th subject is
$$T_i = \exp\left\{\sum_{m=1}^M0.5I(A_{im}=d_{im})-2|X_{i13}|+X_{i15} +\epsilon_{1i}\right\},$$
 and the censoring time is $C_i = \exp\left(0.5|\epsilon_{2i}|+\epsilon_{3i}+\epsilon_{4i}+c_0\right)$, where $\epsilon_{1i}$ is generated from a normal distribution $\mathcal{N}(0,1)$, $\epsilon_{ri}$ ($r=2,3,4$) are generated from continuous uniform distribution $U(0,1)$, and $c_0$ is a constant added to control the censoring rate.

In Example 4, we include non-linear classification functions, and intentionally add some outliers into the samples to evaluate the robustness of the proposed method. {Although all the methods could be affected by the outliers,} the proposed method still performs better
than the
other methods. According to Table \ref{stage4table}, the proposed method outperforms the subgroup and the censored Q-learning methods with improvement rates of $19.4\%$ and $130.4\%$, respectively, for $t_g=2.1$.
{Note that the censored Q-learning method performs much worse in this example than in other examples in terms of the estimated conditional survival probabilities, which
indicates that the censored Q-learning method is not appropriate for data based on non-linear classification functions.
This may be because the censored Q-learning method
is not robust to the non-linear classification functions and the outliers.}

\begin{table}  [H]
	\begin{center}
		\caption{{{Example 4}: The estimated conditional survival probability $\bar{S}^{\widehat{\bm{\mathcal{D}}}}(t_g)$ with standard deviations (SD) in parentheses based on 150 simulations. ``Imp-rate'' represents the improvement rate of the proposed method compared to other methods.}} \label{stage4table}
		\renewcommand{\arraystretch}{1.2} \tabcolsep 0.1in
		\begin{tabular}{cccccccccccccccccc}  \hline  \hline
			Stage & $t_g$ & Censoring Rate  &Method & $\bar{S}^{\widehat{\bm{\mathcal{D}}}}(t_g)$(SD) & Imp-rate \\ \hline
3 & 1.4 & 72\% & Proposed	   &   \textbf{0.752}(0.071)  & --\\
                                                   &   &   &Qlearning & 0.384(0.101) & 96.0\% \\
		        &    &  &Subgroup &  0.679(0.096) & 10.9\% \\

5 & 2.1 & 72\% & Proposed	   &   \textbf{0.659}(0.165) & -- \\
                                                   &   &   &Qlearning & 0.286(0.173)  & 130.4\% \\
		        &    &  &Subgroup &  0.552(0.185) & 19.4\% \\
			\hline
		\end{tabular}
	\end{center}
\end{table}

%

\begin{figure} [H]

   \begin{minipage}{0.4\textwidth}
     \centering
     \includegraphics[width=1\linewidth]{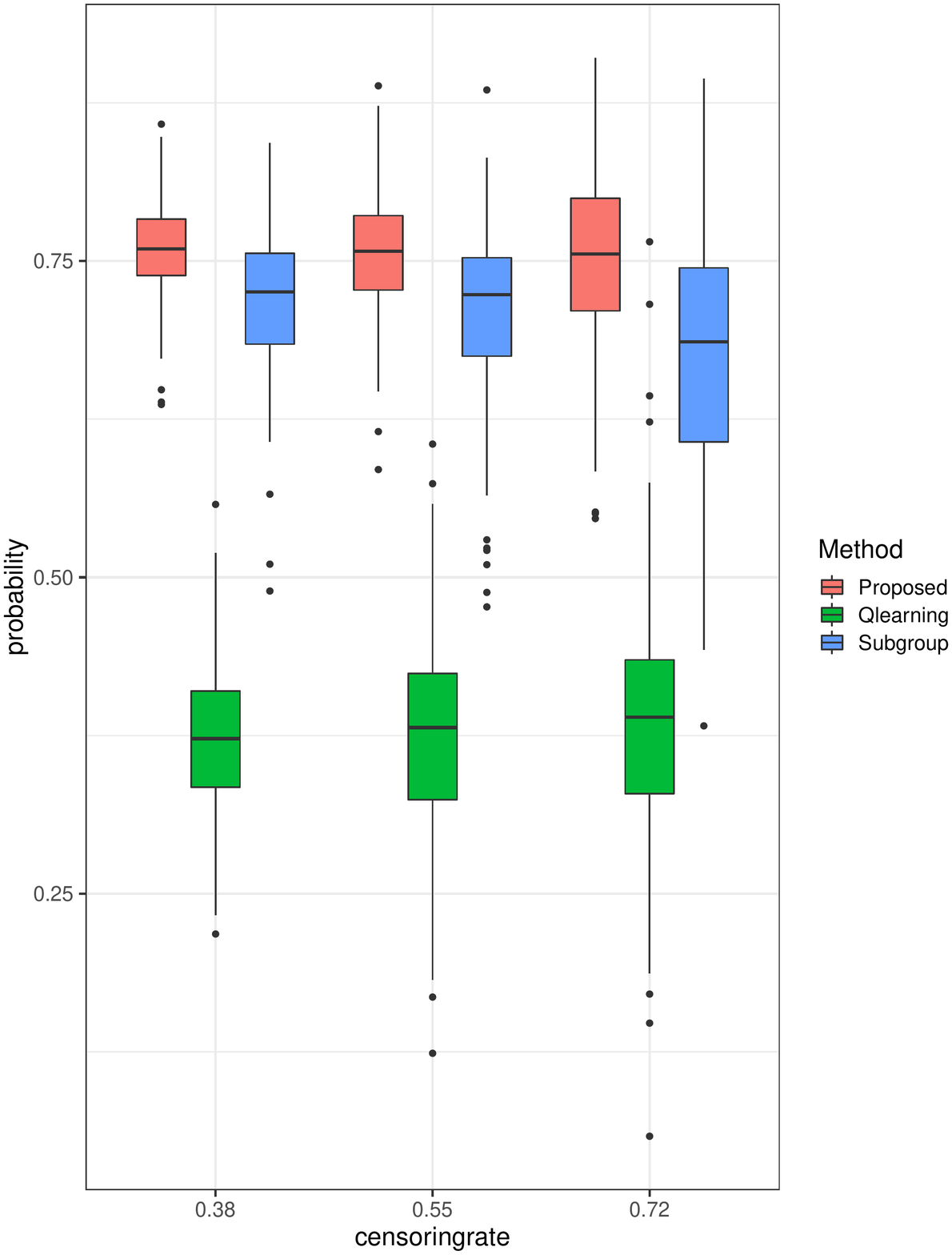}
     \caption{{Example 4}: {Boxplots of the estimated conditional survival probabilities at the third stage for different censoring rate}}\label{Fig:stage4test3}
   \end{minipage}\hfill
   \begin{minipage}{0.4\textwidth}
     \centering
     \includegraphics[width=1\linewidth]{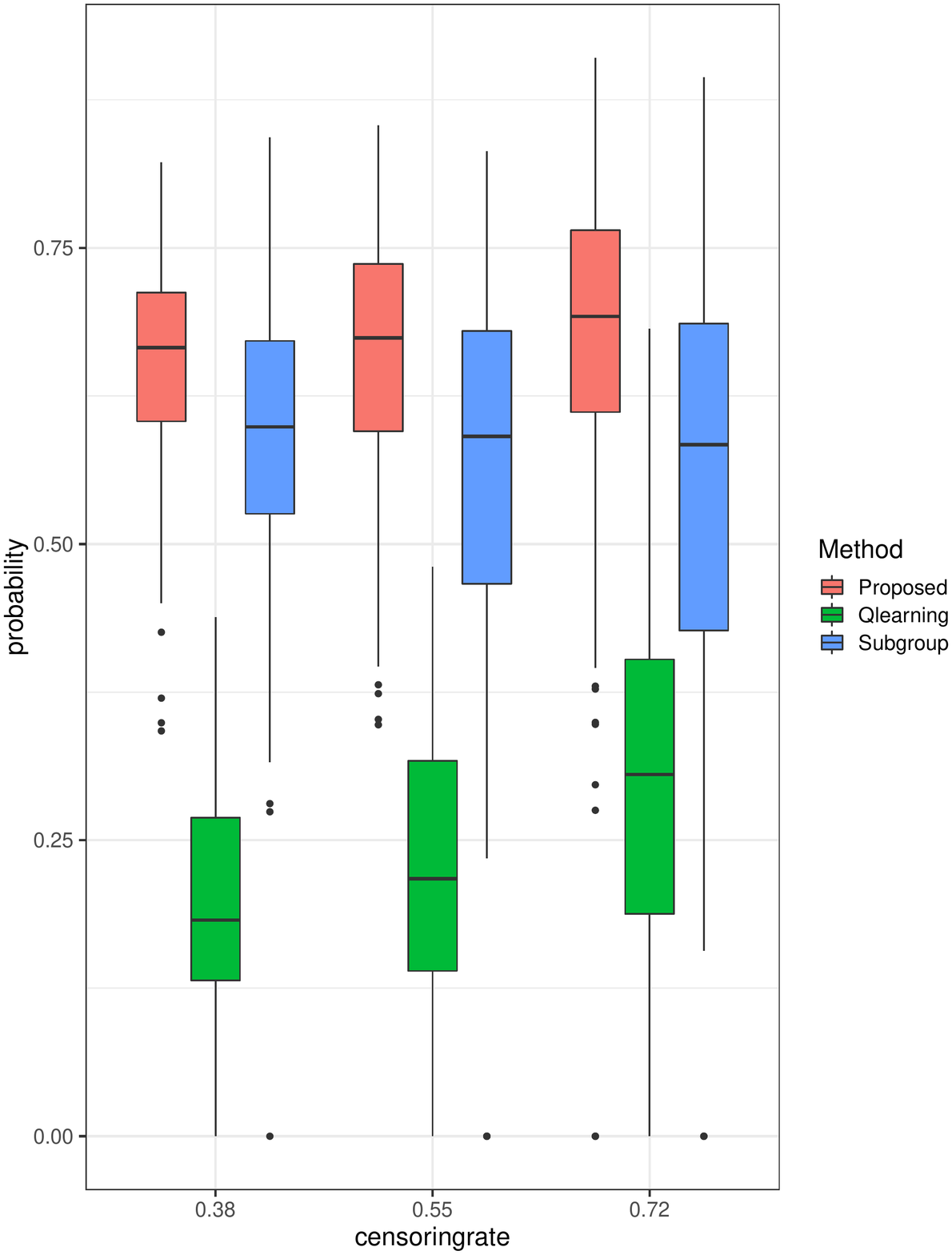}
     \caption{{Example 4}: {Boxplots of the estimated conditional survival probabilities at the fifth stage for different censoring rate}}\label{Fig:stage4test5}
   \end{minipage}
\end{figure}

In summary, the results from the four examples indicate that the proposed method produces better predictions than the competing methods. In addition, Figures \ref{Fig:stage1test3}-\ref{Fig:stage4test5} show that the improvement of the performance from the proposed method increases as the censoring rate increases. This implies that the proposed method can achieve better numerical performance and estimate the optimal dynamic treatment regime more accurately under a high censoring rate. One possible reason is that the proposed method utilizes more information by incorporating patients' information from all stages instead of from only one single stage. For the case of a single decision point, model-based indirect methods such as the censored Q-learning may be efficient if the posited regression model is correctly specified. 
However, in the multistage setting, the verification of true model specification for the survival time might be infeasible for all stages, and thus the censored Q-learning is likely to fail due to model misspecification. Instead, the proposed method circumvents this problem by avoiding modeling the relationship between the survival time and covariates, and estimates the optimal dynamic treatment regime directly.

\section{Real Data Examples}\label{sec:7}

In this section, we apply the proposed method to the Framingham heart study data and acquired immunodeficiency syndrome (AIDS) clinical data for assessing its performance. The proposed method is implemented with linear classification functions.
For comparison, we consider the censored Q-learning method \citep{goldberg2012q} and the subgroup identification method \citep{huling2018subgroup} with one-versus-rest extension, the same as in Section \ref{sec:6}.

\subsection{Application to Framingham Heart Study Data}\label{real1}

The Framingham heart study, established in $1948$, is the first longitudinal prospective large-scale cohort to study cardiovascular disease in the United States. The epidemiological and genetic risk factors for cardiovascular disease are investigated by monitoring original cohort participants. According to \cite{tsao2015cohort}, up to $32$ examinations are performed for each individual biannually during the $65$ years of follow-up. Here we only utilize participants' information from the second visit to the sixth visit, where each visit represents one stage, and different participants can have different numbers of visits. Typically, an antihypertensive treatment will be recommended to individual participants based on the level of blood pressure. {Nevertheless, the accuracy of recommendation  may be potentially improved if we take more information about participants into account, so that the overall survival probability of cardiovascular disease could be improved.}
In this study, we aim to maximize the overall survival probability of cardiovascular disease by customizing sequential decision rules which inform each participant whether an antihypertensive treatment should be taken to balance the blood pressure level at each decision point.

Specifically, we obtain participants' observed survival time ($Y_i$), death reviewed status ($\Delta_i$), visit time, and whether to take the antihypertensive treatment ($A_{im}=0$ or $1$) at each visit $m$ from the data.
In our analysis, we incorporate the following information as covariates ($\bm{X}_{im}$) collected at each visit: age, cholesterol, high-density lipoprotein, diastolic blood pressure, and the presence of diabetes.
There are a total of $3939$ participants who have completely observed covariates at each visit before $Y_i$,
	that is, $n=3939$. The death event rate is $22.9\%$ by the end of the study. We randomly select 1000 participants as a training set and the other 1000 participants as a testing set for 50 iterations.
We consider four target time-points $t_g=4775$, $5946$, $7376$ and $8539$, and calculate the estimated conditional survival probability at each $t_g$ for each testing set given the recommended sequential decision rules estimated on the corresponding training set.}

The average values of the estimated conditional survival probabilities and the corresponding boxplots are presented in Table \ref{realdata_results_test} and Figure \ref{realtest}, respectively. Table \ref{realdata_results_test} and Figure \ref{realtest} show that the proposed method achieves the highest survival probabilities at all different {target time-points}. The improvements of the proposed method compared to the censored Q-learning method and subgroup method are both more than $5\%$. Notice that as the number of stages increases, the proposed method improves even more than the competing methods. This {is possibly due to the fact} that {the proposed method integrates information from all stages into one single algorithm, which is more effective than the methods
only considering information from one single stage information {at a time}}.

\begin{table}  [H]
	\begin{center}
		\caption{
The estimated conditional survival probability $\bar{S}^{\widehat{\bm{\mathcal{D}}}}(t_g)$ with standard deviations (SD) in parentheses based on 50 simulations. ``Imp-rate'' represents the improvement rate of the proposed method compared to other methods.
}  \label{realdata_results_test}
		\vspace{-0.1cm}
		\renewcommand{\arraystretch}{1.1} \tabcolsep 0.2in 
		\begin{tabular}{cccccccccccccc}  \hline  \hline
Stage  & $t_g$   & Method    &	$\bar{S}^{\widehat{\bm{\mathcal{D}}}}(t_g)$(SD) & Imp-rate   \\ \hline
3       &4775  &Proposed  & \textbf{0.988}(0.002) & -- \\
       &        &Qlearning & 0.881(0.012)   & 12.1\%  \\
       &        &Subgroup  & 0.931(0.006)   & 6.1\%  \\
4      &5946  & Proposed  & \textbf{0.940}(0.014) & --  \\
       &        &Qlearning & 0.888(0.008) & 5.9\%   \\
       &        &Subgroup  & 0.895(0.009) & 5.0\%  \\
5       &7376  &Proposed  & \textbf{0.911}(0.024) & --  \\
       &        &Qlearning & 0.684(0.062) & 33.2\% \\
       &        &Subgroup  & 0.790(0.012) & 15.3\% \\
6      & 8539   & Proposed & \textbf{0.802}(0.047) & -- \\
            &   & Qlearning & 0.533(0.038) & 50.5\% \\
            &   & Subgroup & 0.685(0.019) & 17.1\%  \\
\hline
		\end{tabular}
	\end{center}
\end{table}
\vspace{-0.9cm}
\begin{figure} [H]
     \centering
     \scalebox{0.37}[0.37]{\includegraphics{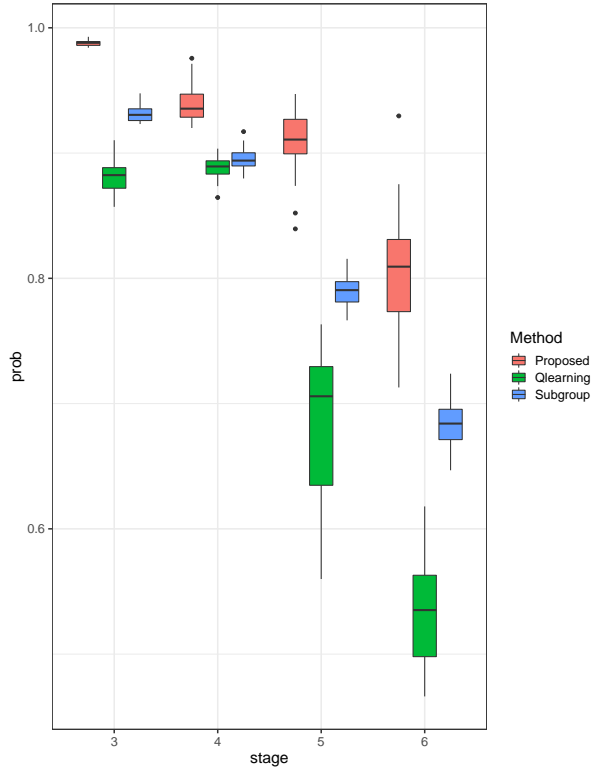}}
     \caption{Boxplot of the estimated conditional survival probabilities based on 50 simulation runs.}\label{realtest}
\end{figure}

\subsection{Application to AIDS Clinical Data}

In this subsection, we apply the proposed method to data from the AIDS Clinical Trials Group (ACTG) 175 \citep{Hammer1996}. ACTG 175 is a double-blind and randomized clinical trial, {which uses the counts of CD4$^+$ T cells of patients to compare monotherapies and combination therapies of HIV infection.}
In this trial, 2139 patients ($n=2139$) with HIV infection were randomly assigned to four treatment groups ($K=4$) with the same probability: zidovudine (ZDV) monotherapy, ZDV and didanosine (ddI), ZDV and zalcitabine (ZAL), and ddI monotherapy.
We obtain information regarding
whether events happen or not for a patient ($\Delta_i$), and the observed time ($Y_i$) when patients are censored or events occur. Here, an event refers to the patient' death, a decline of CD4$^+$ T cell counts being at least $50$, or an event indicating progression to AIDS.
In addition to the treatment and event time, we consider $12$ clinical covariates {($p=12$)} in our analysis, which are also included in \citep{FanCaiyun2017} and \citep{qi2018}. Five of the 12 covariates are continuous: weight, CD4$^+$ T cells counts at baseline, CD8 count at baseline, age, and Karnofsky score, {where Karnofsky score is a way to rate a person's ability to perform activities of daily living, and ranges from 0 to 100, with a higher score indicating that a person is more able to perform daily activities.}
The remaining seven covariates are binary: gender (0=female, 1=male), race (0=white, 1=non-white), homosexual activity (0 = no, 1 = yes), history of intravenous drug use (0=no, 1=yes), symptomatic status (0=asymptomatic, 1=symptomatic), antiretroviral history (0=naive, 1=experienced), and hemophilia (0=no, 1=yes).

This study only involves one decision point (one stage), but with multicategory treatments which contain four different treatment choices.
Similarly as in Subsection \ref{real1}, we randomly select 1000 participants as a training set and the other 1000 participants as a testing set for 50 iterations.
We specify three target time-points $t_g=861$, $984$ and $1169$, and calculate the estimated conditional survival probability at each $t_g$ for each testing set given the recommended sequential decision rules estimated on the corresponding training set.

Table \ref{realdata_results_test2} and Figure \ref{realtest2} show that the proposed method achieves the highest survival probabilities at all target time-points. Notice that as the $t_g$ increases, the proposed method has improved more compared to the competing methods.
}{Although there is} only one stage for treatment decision, the higher survival probabilities {of the proposed method} still indicate that patients would receive {more} long-term benefits if the proposed strategy is {applied}.

  \begin{table}  [H]
	\begin{center}
		\caption{
The estimated conditional survival probability $\bar{S}^{\widehat{\bm{\mathcal{D}}}}(t_g)$ with standard deviations (SD) in parentheses based on 50 simulations. ``Imp-rate'' represents the improvement rate of the proposed method compared to other methods.
}  \label{realdata_results_test2}
		\vspace{-0.1cm}
		\renewcommand{\arraystretch}{1.1} \tabcolsep 0.2in 
		\begin{tabular}{cccccccccccccc}  \hline  \hline
 $t_g$     &Method    &	$\bar{S}^{\widehat{\bm{\mathcal{D}}}}(t_g)$(SD) & Imp-rate   \\ \hline
861    &Proposed  & \textbf{0.779}(0.050) & --  \\
         &Qlearning & 0.774(0.046) & 0.65\%   \\
         &Subgroup  & 0.753(0.071) & 3.45\%  \\
984    &Proposed  & \textbf{0.823}(0.052) & -- \\
         &Qlearning & 0.709(0.050)   & 1.97\%  \\
         &Subgroup  & 0.702(0.069)   & 2.99\%  \\
1169     &Proposed  & \textbf{0.695}(0.055)  &  --  \\
         &Qlearning & 0.669(0.057)  & 3.89\%  \\
         &Subgroup  & 0.671(0.069)  & 3.58\%  \\
\hline
		\end{tabular}
	\end{center}
\end{table}

\vspace{-0.9cm}

\begin{figure} [H]
     \centering
     \scalebox{0.37}[0.37]{\includegraphics{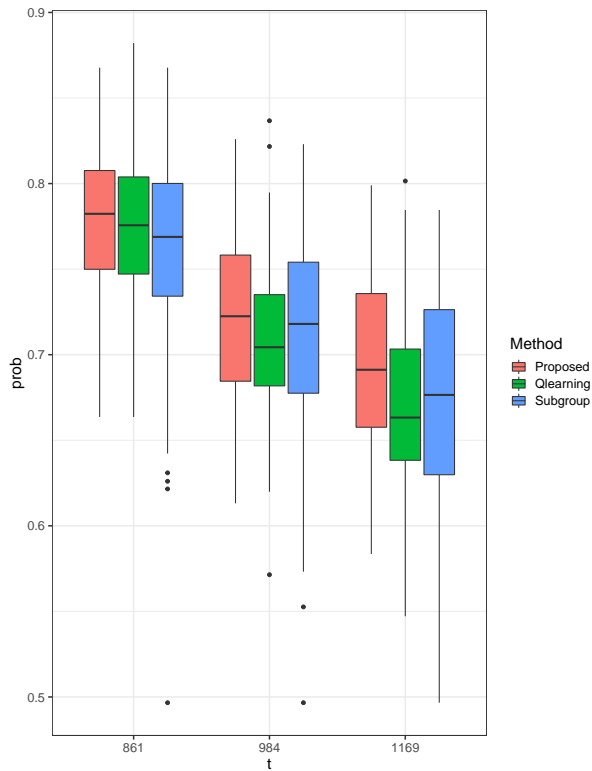}}
     \caption{Boxplot of the estimated conditional survival probabilities based on 50 simulation runs.}\label{realtest2}
\end{figure}

\section{Discussion}\label{sec:8}

In this paper, we propose a novel weighted Kaplan-Meier estimator to estimate the survival function for patients following a specific {DTR}, and propose an objective function based on this weighted Kaplan-Meier estimator under an angle-based framework for multicategory treatments. We estimate the optimal DTR through maximizing the proposed objective function.
In this way, we formulate the estimations of decision rules at multiple stages
as an unconstrained optimization problem.
Therefore, the proposed method is more computationally efficient than the multicategory methods with the sum-to-zero constraint.
	Also, the proposed method
	avoids a potential model-misspecification problem which commonly arises in
	fitting the regression model to one stage based on 
	estimations of regression models at other stages.
Moreover, the proposed method can be easily extended to incorporate some other multicategory classification approaches due to its flexibility.

Several improvements and extensions are worth exploring in the future. Although the proposed method is implemented for a linear classification function based on a linear combination of prior information in this paper, we can also extend our method to estimate non-linear classification functions by applying Gaussian kernel learning classifiers.
In addition, the goal of the proposed method is to search for the optimal DTR for maximizing conditional survival probability, which can be potentially generalized to obtain DTRs that maximize other outcomes of interests, e.g., mean residual life and median survival time.

Additionally, developing statistical inference for prediction is also useful in practice, e.g., generalizing our method to quantify uncertainty via confidence interval and hypothesis testing.
Moreover, the proposed method is based on an inverse probability weighting procedure, which is potentially less efficient since it only utilizes information of patients who received treatments following a given DTR.
To incorporate information of patients who do not follow a given DTR, a potential generalization of the proposed method could be based on augmented inverse probability weighting \citep{zhang2013robust}. 
Furthermore, it is important to extend the proposed method for continuous treatments, such as drug dosages, to
improve clinical practice and enhance personalized medicine treatment for patients.

\newpage

\bibliographystyle{agsm}
{\footnotesize
\bibliography{my_bib}}


\end{document}